\documentstyle[epsf]{mn}

\newif\ifAMStwofonts

\newcommand{\hi}{H\,{\sc i}}
\newcommand{\hii}{H\,{\sc ii}}

\newcommand{\etal}{et al.}

\newcommand{\Ein}{{\em Einstein}}

\newcommand{\Ros}{{\em ROSAT}}
\newcommand{\ros}{{\em ROSAT}}
\newcommand{\ltsim}{\raisebox{-1mm}{$\stackrel{<}{\sim}$}}
\newcommand{\lapx}{\raisebox{-1mm}{$\stackrel{<}{\sim}$}}

\newcommand{\eg}{e.g. }
\newcommand{\ie}{i.e. }
\newcommand{\etc}{etc.}
\def\arcm{\hbox{$^\prime$}}
\def\arcs{\arcm\hskip -0.1em\arcm}

\title[The X-ray Evolution of Merging Galaxies]
{The X-ray Evolution of Merging Galaxies}

\author[A. M. Read \& T. J. Ponman] 
 {A. M. Read$^{1,2}$ \& T. J. Ponman$^{2}$ \\
$^{1}$ Max-Planck-Institut f\"{u}r extraterrestrische 
Physik, Postfach 1603, D-85740, Garching, Germany  \\
$^{2}$ School of Physics and Astronomy, 
University of Birmingham, Edgbaston, BIRMINGHAM, B15 2TT, U.K.}

\pagerange{\pageref{firstpage}--\pageref{lastpage}}
\def\LaTeX{L\kern-.36em\raise.3ex\hbox{a}\kern-.15em
    T\kern-.1667em\lower.7ex\hbox{E}\kern-.125emX}

\begin{document}

\label{firstpage}

\maketitle

\begin{abstract}

We present here the first study of the X-ray properties of an
evolutionary sample of merging galaxies. Both \Ros\ PSPC and HRI data are
presented for a sample of eight interacting galaxy systems, each
believed to involve a similar encounter between two
spiral discs of approximately equal size. The mergers span a large range
in age, from completely detached to fully merged systems.

A great deal of interesting X-ray structure is seen, and the X-ray
properties of each individual system are discussed in detail. Along the
merging sequence, several trends are evident: in the case of several of
the infrared bright systems, the diffuse emission is very extended, and
appears to arise from material ejected from the galaxies. The onset of
this process seems to occur very soon after the galaxies first encounter
one another, and these ejections soon evolve into distorted flows. More
massive extensions (perhaps involving up to $10^{10}$\,$M_{\odot}$ of hot
gas) are seen at the `ultraluminous' peak of the interaction, as the
galactic nuclei coalesce.

The amplitude of the evolution of the X-ray emission through a merger is
markedly different from that of the infrared and radio emission however.
Although the X-ray luminosity rises and falls along the sequence, the
factor by which the X-ray luminosity increases, relative to the optical,
appears to be only about a tenth of that seen in the far-infrared. This
we believe, may well be linked with the large extensions of hot gas
observed.

The late, relaxed remnants, appear relatively devoid of gas, and possess
an X-ray halo very different from that of typical ellipticals, a problem
for the `merger hypothesis', whereby the merger of two disc galaxies
results in an elliptical galaxy. However, these systems are still
relatively young in terms of total merger lifetime, and they may still
have a few Gyr of evolution to go through, before they resemble typical
elliptical galaxies.

\end{abstract}


\section{Introduction}
\label{sec_intro}

Galaxies were once thought of as `island universes', evolving slowly in
complete isolation. This is now known not to be the case. Galaxies
interact in a variety of ways with their environment, both with satellite
and neighbouring galaxies, and, in the case of galaxies within groups and
clusters, with large masses of tenuous hot gas. Collisions and mergers of
galaxies are now thought to be one of the most dominant evolutionary
mechanisms (Schweizer 1989). Indeed, if events during galaxy formation
are counted, there are probably very few galaxies that were not shaped by
interactions or even outright mergers (Toomre 1977). The position of a
galaxy in Hubble's (1936) morphological sequence may in fact depend
mainly on the number and severity of merger events in its past history.
Pure disc systems, formed from relatively isolated protogalactic gas
clouds, appear at one end of the Hubble sequence, the giant ellipticals,
possibly produced through mergers of similar spirals, appear at the
other, and in between, mergers between galaxies of differing mass produce
galaxies with a wide range of bulge to disc ratios.

This {\em merger hypothesis} discussed above, the idea that elliptical
galaxies might be formed from the merger of two disc galaxies, was first
suggested by Toomre (1977). The conversion of orbital to internal energy
during a close tidal encounter causes the two progenitor systems to sink
together and coalesce violently into a centrally condensed system,
disrupting any pre-existing discs, and largely randomizing the stellar
motions, as shown in many N-body simulations (\eg Barnes 1988). The
problem that ellipticals have central mass densities some $10^{2}-10^{3}$
times higher than normal spirals has been overcome by the discovery that
young merger remnants contain huge central concentrations of molecular
gas (Sanders \etal\ 1986; Sanders \etal\ 1988a; Scoville \& Soifer 1991).
Also, elliptical-like luminosity profiles have been discovered in the
infra-red underlying the disturbed optical morphology of merging galaxies
(Wright \etal\ 1990). There have been many attempts to disprove
the merger hypothesis. Counter-evidence cited includes the fractional
abundances of ellipticals in clusters, the colour-luminosity and
metallicity-luminosity relationships, radial gradients within ellipticals
and with the incidence of globular clusters within ellipticals (see
Hernquist 1993). These criticisms can mostly be overcome by appealing to
a variety of ideas, such as the growth of substructure in hierarchical
cosmogonies, the incomplete nature of violent relaxation, and merger
induced star-formation.

Studies of the evolution of merging galaxies have been performed in many
regions of the electromagnetic spectrum (Joseph \& Wright 1985; Telesco,
Wolstencroft \& Done 1988; Casoli \etal\ 1991; Hibbard \& van Gorkom 1996),
and these are discussed in detail in both Section~\ref{sec_sample} and
Section~\ref{sec_discussion}. One region of the electromagnetic spectrum that
has been neglected so far though, has been the X-ray.

The \Ros\ X-ray telescope (XRT), with the Position Sensitive Proportional
Counter (PSPC) (Pfeffermann \etal\ 1986) at the focal plane, offers three
important improvements over previous X-ray imaging instruments, such as
the \Ein\ IPC. First of all, its spatial resolution is very much
improved, the 90\% enclosed energy radius at 1\,keV being $27''$
(Hasinger \etal\ 1992). Secondly, the spectral resolution of the PSPC is
significantly better ($\Delta E/E \sim 0.4$ FWHM at 1\,keV) than earlier
X-ray imaging instruments and this allows the derivation of characteristic
source and diffuse emission temperatures. Finally, the PSPC internal
background is very low ($\sim3\times10^{-5}$\,ct s$^{-1}$ arcmin$^{-2}$;
Snowden \etal\ 1994), thus allowing the mapping of low surface brightness
emission. The High Resolution Imager (HRI) on the other hand, because of
its excellent spatial resolution ($\approx5''$) and relative insensitivity to
diffuse emission, is an ideal instrument for further investigation into
the point source populations.

Furthermore, as \Ros's energy band is relatively soft compared to that of
the \Ein\ IPC (and to other previous instruments), the hot halos now known
to exist around many nearby normal galaxies (\eg Bregman \& Pildis 1994; Wang
\etal\ 1995; Read, Ponman \& Strickland 1997), may be observable within
these more distant, though more active systems. Indeed as shall be
discussed later, a hot halo is known to exist around the merging galaxy
pair, the {\em Antennae} (NGC~4038/9) (Read, Ponman \& Wolstencroft
1995).

In the \Ros\ band, it appears that the X-ray emission from normal spiral
galaxies is made up of a complex mix of sources. Both stellar sources
such as low- and high-mass X-ray binaries, cataclysmic variables and
normal main sequence stars, and interstellar sources such as supernova
remnants (SNRs) and the hot phases of the ISM, contribute to the total
X-ray emission. The merger of two similar spiral galaxies may lead to
additional sources of X-ray emission. The advances in instrumentation
discussed above should allow a substantial improvement in our knowledge
of the complex X-ray properties of interacting galaxies over what was
possible with \Ein\ (see Fabbiano 1989 and references therein). The
variation in X-ray emission with merger stage and position can be
compared with corresponding multiwavelength results to reveal new clues
as to how the tidal forces trigger star formation, and to constrain the
nature and duration of the activity which ensues.

Studies of the nearest, and therefore best resolved, interacting systems
are especially important, and many of the most attractive targets have
been observed by \Ros. We have undertaken a programme of \Ros\ PSPC and
HRI observations of a carefully chosen chronological sequence of merging
galaxies in order to study the evolution of their X-ray properties
through the merging process. We have combined our own \Ros\ data with
other archival \Ros\ data and with individual studies published recently.
Where possible, we have attempted to analyse these systems in a uniform
manner. In particular we have carefully separated point sources from
what appears to be diffuse X-ray emission, and have determined the spectral
properties of each.

In the present paper we describe the analysis we have performed, and
present the basic results for each of our sample systems in turn. In some
cases, we are able to compare with previous authors' results from the
same data, but in many instances our analysis is the first reported \Ros\
analysis, and in some cases it represents the first picture of the X-ray
emission from an imaging telescope.

The plan of this paper is as follows. After some comments
(Section~\ref{sec_sample}) on the selection of the sample,
Section~\ref{sec_obs_data} describes the observations and the data
reduction methods. The results for each galaxy are presented in
Section~\ref{sec_results}, together with notes on the individual systems.
In Section~\ref{sec_discussion}, we comment on the range of
galaxy properties which emerges, and discuss the X-ray evolution of the
sample as a whole. Finally, in Section~\ref{sec_conclusions} we present
our conclusions.

\section{The merging galaxy sample}
\label{sec_sample}

The basic problem with a chronological study is to establish the ordering
of the sample. The most famous chronological sequence of merging galaxies
is undoubtedly the `Toomre sequence' (Toomre 1977), containing the 11
best examples of ongoing mergers from the New General Catalogue of
Nebulae and Clusters of Stars (NGC). These systems appear, when ordered
chronologically, to range from well separated, strongly interacting
systems, apparently destined to eventually merge, through systems in the
throes of merging, to well merged systems. This sequence has formed the
basis for other evolutionary studies made at other wavelengths, as
discussed below, and indeed, forms the basis for the selection of the
present sample. 

Relevant studies at other wavelengths were, until very recently, 
rare. Joseph and Wright (1985) studied highly disturbed systems in which
the two disc galaxies had lost their individual identities and appear as
a single coalesced object, the primary morphological indication of which
is the presence of the two tidal tails. Dynamical simulations (Toomre and
Toomre 1972) suggested that the tails will persist for $\sim 10^{9}$ years,
and the faintness of these tails, together with the degree of
coalescence, were used as indicators of relative merger age. Comparing
the infrared luminosities of these galaxies they found that the two
youngest and two oldest systems were less luminous than the four
`middle-aged' systems, suggesting that merging galaxies undergo a phase
of starburst activity of characteristic luminosity $\sim 10^{12}
L_{\odot}$. Another useful indicator of the progress of a starburst was
found by Telesco, Wolstencroft \& Done (1988); for comparably sized
interacting galaxies, the dust colour temperature and the interaction
strength is greatest for the pairs with the smallest separation. Furthermore, 
Casoli \etal\ (1991), while studying the molecular gas content within a 
series of merging systems, discovered that the far-infrared colour
temperature, the $L_{FIR}/L_{B}$ ratio and the $L_{FIR}/M$(H$_{2}$) ratio all 
evolve
along the sequence, increasing in the early stages of the collision, reaching
a climax, and then decreasing to values typical of ellipticals. 

Much more recently, Hibbard \& van Gorkom (1996) have presented \hi,
\hii, and R-band observations of an evolutionary sequence of merging
systems, again taken primarily from the Toomre sequence. Their five systems,
four of which appear in the Toomre sequence, and three of which appear in
this present paper, span a large range in merger stage, unlike the work of
Joseph \& Wright (1985) discussed above, where only middle stage mergers
were considered. Much of Hibbard \& van Gorkom's (1996) work is very relevant
to the present paper, and many of their results are used in our discussion
(Section~\ref{sec_discussion}). They find that, moving from early to
late-stage mergers, a larger fraction of the neutral hydrogen is found outside
of the optical confines of the system, and in the last stage, all of the \hi\
is found in the tails, with none in the central regions. They suggest that the
interaction-induced relaxation causes most of the atomic gas within the
original discs to compress, condense, form stars and/or shock-heat to X-ray
temperatures, any residual atomic gas remaining in the remnant being heated
via the resultant starburst.

In order to make the establishment of a reliable chronological ordering as
easy as possible, the systems were carefully chosen on the basis of both close
attention to the work of the previous authors' described above,
especially Toomre's (1977) sequence, and the following important criteria:

\begin{itemize}

\item All the systems contain, or appear to have evolved from, two spiral 
galaxies of fairly equal mass. 

\item A wide range of merger stages, wider in fact than the Toomre
sequence, from completely detached to fully merged 
systems, is covered. 

\item All the systems are infrared and radio bright, indicating the presence 
of unusual star-formation activity. 

\item Low absorbing columns ($1.2-5.2\times10^{20}$\,cm$^{-2}$) exist in the
directions of all these systems, maximizing the sensitivity to soft X-ray
emission. 

\item All the systems are large enough for worthwhile spatial resolution to be 
achieved with the PSPC, and for detailed mapping of point sources with the 
HRI. 

\item Much additional multiwavelength information is available for all 
these systems. 

\end{itemize}

Optical morphology (especially the appearance of tidal tails, and the
proximity of the two nuclei) has been used as the primary measure of merger
stage, but this has been supplemented with the following additional indicators.

Toomre \& Toomre (1972) have shown that proper tail construction requires 
that the perturbing mass be at least comparable to the perturbed, and that 
therefore, major tidal tails should protrude from both of two interacting discs
of roughly equal dimensions as in NGC~4676 (the {\em Mice}) and NGC~4038/9 (the
{\em Antennae}). For systems where the central parts of the galaxy have not
yet merged, tail length seems the best indicator of age, increasing with time
(Toomre \& Toomre 1972), although tail length also increases with the
closeness of the encounter. Once the nuclei of the galaxies have merged, the
tails begin to diffuse away, and so using the morphology of the systems to
order them chronologically comes down to a combination of tail length, tail
faintness and degree of coalescence of the nuclei. 

Listed below, in chronological order, are the eight systems within the sample, 
selected on the basis of the above considerations. 
Optical pictures of the 
eight systems are shown in figure~\ref{mergers_pics}. All are taken from the 
digitised sky survey, the southern systems from the U.K. Schmidt plates and the 
northern systems from the Palomar Schmidt plates.

{\bf Arp~270} (NGC~3395/6) consists of two low IR-flux galaxies, obviously
involved in some sort of interaction, though the lack of 
tidal tails is suggestive of them not having properly encountered
one another yet. This system is in fact a pre-Toomre sequence system.

{\bf Arp~242}, the {\em Mice} (NGC~4676), occurs second in the Toomre
sequence and, though it appears rather similar to Arp~270, the presence
of tidal tails and an obvious bridge between the two galaxies, together
with an increase in IR activity, indicates that the galaxies have begun
to interact, and have passed each other.

{\bf NGC~4038/9}, the {\em Antennae} (Arp 244), is a classic example of
an interacting system with two equally long tails and distinct central
masses. Although it occurs first in the Toomre sequence, CO (Stanford
\etal\ 1990), radio (Hummel \& van der Hulst 1986) and X-ray
observations (Read \etal\ 1995) indicate that the discs have begun to
merge, and the {\em Antennae} therefore should really be placed sometime
after the {\em Mice}.

{\bf NGC~520} (Arp 157), seventh in the Toomre sequence, and classified
as an intermediate-stage merger by Hibbard \& van Gorkom (1996), is as
radio and infrared bright as the {\em Antennae}, and has two smaller
tails as well as two nuclei and two velocity systems in its spectra,
indicative of a young merger.

{\bf Arp220}, one of the superluminous IRAS galaxies, is a prototypical
merger. It is the brightest of the sample across the whole of the
electromagnetic spectrum (except for the optical). Huge H$\alpha$
filaments exist (Heckman, Armus \& Miley 1990), suggestive of vigorous
starburst activity. Two distinct (though very close) nuclei are visible in the
infrared (Majewski \etal\ 1993).

{\bf NGC~2623} (Arp 243), appearing eighth in the Toomre sequence, is, 
like Arp~220, a superluminous IRAS galaxy, and is also very bright in the
radio.  Very long tails are visible but the central masses have become 
indistinguishable. Only one true nucleus is thought to exist. 

{\bf NGC~7252} (Arp 226), the {\em atoms for peace} galaxy, is the
prototypical merger remnant, and is the last system within the Toomre
sequence. It has two equally long, very faint tails emanating from a
disturbed, double motion spheroidal body whose mean light distribution
closely follows the de Vaucouleurs $r^{1/4}$ law, typical of ellipticals.

{\bf AM~1146-270}, beyond the end of the Toomre sequence, is a spheroidal
galaxy with a small tail-like structure, appears to be the site of a
great deal of recent star formation, hence the many blue knots
surrounding its nucleus.

\begin{figure*}
\vspace{140mm}
\caption[The merging galaxy sample.]{\small The merging galaxy sample: A)
Arp~270, B) Arp~242, C) NGC~4038/9, D) NGC~520, E) Arp~220, F) NGC~2623,
G) NGC~7252, H) AM~1146-270}
\label{mergers_pics}
\end{figure*}

Table~\ref{table_sample} lists the sample systems together with the
following basic properties. Distances are, where possible, taken from
Tully (1988). These are based on $H_{0}=75$\,km s$^{-1}$ Mpc$^{-1}$, and
assume that the Galaxy is retarded by 300\,km s$^{-1}$ from universal
expansion by the mass of the Virgo cluster. For the more distant
galaxies, distance values are either taken from Condon \etal\ (1990),
where the same $H_{0}$ and Virgocentric flow correction is used, or are
calculated from the group-flow-corrected radial velocities given in
the Second Reference Catalog of Bright Galaxies (RCBG) (de Vaucouleurs,
de Vaucouleurs \& Corwin 1976), using a $H_{0}$ of 75\,km s$^{-1}$
Mpc$^{-1}$. AM~1146-270's radial velocity is taken from Sekiguchi \& Wolstencroft (1993). 

Optical (B) luminosities for all the systems in the sample are calculated
as in Tully (1988), \ie as a re-expression of the absolute magnitude,
itself following from the blue apparent magnitude $B_{T}$,  and the
distance $D$:

\begin{displaymath}
\log{L_{B}}(L_{\odot}) = 12.192 - 0.4B_{T} + 2 \log{D}.
\end{displaymath}

Blue apparent magnitudes are, again where possible, taken from Tully
(1988). Values for the more distant systems are taken from RCBG (de Vaucouleurs,
de Vaucouleurs \& Corwin 1976) and from NGC 2000.0 (Dreyer 1988) (the
value for AM~1146-270 is taken from Smith \& Hintzen 1991)). 

FIR luminosities are calculated from IRAS 60 and 100\,$\mu m$
fluxes (taken from the IRAS Point Source Catalogue) using the expression 

\begin{displaymath}
L_{FIR} = 3.65\times10^{5}\left[2.58 S_{60\mu m} + S_{100\mu m}\right]
D^{2} L_{\odot},
\end{displaymath}

(\eg Devereux \& Eales 1989). Here $D$ is the distance in Mpc and
$S_{60\mu m}$ and $S_{100\mu m}$ are the IRAS 60 and 100\,$\mu m$ fluxes
(in Janskys). Also given in table~\ref{table_sample} are the infrared to
blue luminosity ratio, $L_{FIR}/L_{B}$, and the dust colour temperature,
$S_{60}/S_{100}$. These two infrared indicators are the two suggested by
previous studies (Joseph \& Wright 1985; Telesco \etal\ 1988) to be
worthwhile indicators of starburst activity, and it can be seen that they
rise and fall along the chronological sequence in just the way one might expect if a burst of
star formation were triggered during a merger. Finally, included also in
Table~\ref{table_results} are the radio luminosities from Condon \etal's (1990) 1.49\,GHz
atlas of the {\em IRAS} bright galaxy sample, supplemented in the cases of Arp~242 and
NGC~7252 (upper limit), with the luminosities (converted to H$_{0}=75$km s$^{-1}$ Mpc$^{-1}$)
from Heckman's (1983) study.

\begin{table*}
\begin{center}
\begin{tabular}{|l|l|c|c|c|c|c|c|}	\hline
System &Other names & Distance & $\log{L_{B}}$ & $\log{L_{FIR}}$ &
	$L_{FIR}/L_{B}$ & $S_{60}/S_{100}$ & $\log{L_{rad}}$ \\ 
       &            & (Mpc)    &  (erg s$^{-1}$) & (erg s$^{-1}$) & 
                        &      & W\,hz$^{-1}$             \\ \hline
Arp~270    &NGC~3395/6& 28  & 44.08 & 43.61 & 0.335 & 0.486 & 21.93 \\
Arp~242    &NGC~4676  & 88  & 44.06 & 44.09 & 1.069 & 0.534 & 22.18 \\
NGC~4038/9 &Arp~244   & 25  & 44.19 & 44.20 & 1.026 & 0.516 & 22.64 \\
NGC~520    &Arp~157   & 28  & 43.87 & 44.15 & 1.900 & 0.651 & 22.24 \\
Arp~220    &UGC~9913  & 76  & 43.96 & 45.52 & 36.04 & 0.884 & 23.39 \\
NGC~2623   &Arp~243   & 78  & 43.78 & 44.87 & 12.28 & 0.862 & 22.85 \\
NGC~7252   &Arp~226   & 63  & 44.28 & 44.00 & 0.526 & 0.565 &$<$22.78 \\
AM~1146-270&          & 25  & 42.93 & 42.55 & 0.416 & 0.590 &  -    \\
\hline
\end{tabular}
\caption[]{\small The merging galaxy sample, listed in order of evolutionary
age. Sources for the tabulated parameters are given in the text.} 
\label{table_sample}
\end{center}
\end{table*}

\section{Observations And Data Reduction}
\label{sec_obs_data}

Both PSPC and HRI data have been analysed within this study, and the
methods of data extraction and reduction for both types of data are
described below. All of the HRI data and the PSPC data for three of the
systems (NGC~4038/9, NGC~2623 and AM~1146-270) were obtained through our
own pointed observations. The remaining PSPC datasets were obtained from
the UK \Ros\ Data Archive Centre at the Department of Physics and
Astronomy, Leicester University U.K.

\subsection{PSPC Observations and Data Reduction}
\label{sec_pspc}

All of the PSPC datasets have been analysed in essentially the
same way using the STARLINK {\em ASTERIX} X-ray analysis system. This
method is described below, and in greater detail in Read \etal\ (1997).
Departures from this standard procedure
are described in detail in the notes on the individual
systems (sections~\ref{sec_arp270}$-$\ref{sec_am1146-270}). In every case, the
target system in question lay at the centre of the PSPC field, thus minimizing
the effects of vignetting and blurring.

Once the data were `cleaned' of high background periods ($\approx2-3$\%
of the data), they were binned into a (0.1$-$2.3\,keV) spectral image (or
`data cube') of size approximately twice that of each system. An annulus
situated outside the PSPC central support ring, and with the bright
sources within it removed, was formed from the data, and this background
region was, together with knowledge of the PSPC vignetting function, used
to construct a background model. When this was subtracted from the data
cube, the resultant background-subtracted data cube could be used to form
images in different spectral bands.

PSS (Allan, Ponman \& Jeffries, in preparation), a point source search program
which uses a likelihood technique to search for enhancements above the background,
was used to search for point-like emission within these images. The basic method
of PSS involves a comparison of the input dataset (in this case, the image) with a
model, comprising of a background and a scaled PSF. The positions of the detected
point sources, \ie those with a significance $\geq 4 \sigma$ (except where stated 
in the text), were cross-correlated with a variety of stellar and non-stellar
catalogues, including the SIMBAD catalogue and the \Ein\ and IRAS point source
catalogues.

As in Read \etal\ (1997), we attempted to separate the diffuse from the
source emission by removing data from the background-subtracted data cube
at the position of each PSS source over a circle enclosing 93\% of the
energy of a 0.5\,keV point source (as all the sample systems subtended
small angles, this extraction radius was never greater than about half an
arcminute). The remaining data were then collapsed into a spectrum and
corrected for vignetting effects and exposure time. To account for the
diffuse flux lost in the source removal process, the diffuse spectrum was
renormalised using a `patched' image, where the `holes' left after source
removal were filled by bilinear interpolation.

This method of separating the point source emission from the diffuse
emission did not, in general, work as well as in Read \etal\ (1997). This
is not surprising as the systems in the present study are at much greater
distances than those studied in Read \etal\ (1997), and both limitations
on sensitivity and on resolution have had a more serious impact.

As in Read \etal\ (1997), source spectra were binned directly from the
raw data from circles of the same size as those used in the source
removal procedure above. To obtain true {\em source} spectra however,
{\em both} the local true background flux, obtained by scaling the
background emission from source-free regions of the field using knowledge
of the PSPC vignetting function, {\em and} the local diffuse flux,
calculated by scaling the integrated spectrum of the diffuse emission
with an estimate of the fraction of the diffuse flux underlying each
source, had to be removed from each source spectrum.

So, in summary, three types of spectrum were formed from each dataset:

\begin{itemize} 
\item An {\bf integrated Spectrum} was extracted from a circular
area of diameter slightly greater than the optical diameter of each
system. Although particle and cosmic X-ray backgrounds were removed, no
sources were excluded. Standard spectral models (power law,
bremsstrahlung and Raymond and Smith (1977) hot plasma) were used in the
fitting. Even though these spectra contain emission from everything
around each system's position, including galactic sources, galactic
diffuse emission, and foreground and background sources (both Galactic
and extragalactic), some of the resultant spectral fits are very
acceptable. The net number of counts, the $\chi^{2}$ values, number of
degrees of freedom (n.d.o.f.), and the integrated X-ray luminosities
(0.1$-$2.0\,keV) inferred from the spectral fitting, are shown in each of
Tables\,\ref{results_arp270} to \ref{results_ngc7252}.

\item A {\bf Diffuse spectrum} was extracted, using the method described
above, from an area the same size as that used for the
integrated spectrum. The correctly normalized (0.1$-$2.0\,keV) X-ray
luminosities of each system's diffuse emission are tabulated in
Tables\,\ref{results_arp270} to \ref{results_ngc7252}, along with the net
diffuse counts and the results of the spectral fits (using again the same
models as those used above). Because of low count rates, $\chi^{2}$ fitting
could only be used in the case of the {\em Antennae}. In all other cases, a 
maximum likelihood criterion, which allows for the Poissonian nature of the 
data, had to be used. Unfortunately, when using likelihood fitting, the 
absolute value of the statistic is not informative (and hence we have not
given it). The Cash statistic (Cash 1979) can be used however, to compare the
relative quality of different fits, and to derive confidence intervals for
model parameters, since {\em differences} in the value of the Cash statistic
from one model to another, are distributed in the same way as $\chi^{2}$.
Furthermore, only the freezing of certain parameters at sensible values led to
acceptable fits in a few cases (these are indicated by a bracketted
`F'). All these results are described in detail in Sections~\ref{sec_arp270}
to \ref{sec_am1146-270}. Because of the large distances involved, many of
these systems' diffuse emission spectra will be complicated, involving some
genuinely diffuse gaseous emission, and some contribution from both unresolved
point sources and from `stumps' left in the subtraction of the bright sources.
Hence, we attempted, where possible, to fit a 2-component model to the diffuse
spectrum, comprising of a Raymond and Smith hot plasma (representing the truly
diffuse gas) and a highly absorbed hot (10\,keV) component (representing
unresolved sources). These results are also described in more detail later in
the notes on the individual systems.

\item {\bf Individual source spectra} were extracted as described above.
Sources known not to be connected with the galaxy were excluded from the
analysis, as were sources that lay outside both the optical emission and
any diffuse X-ray emission. The results of the spectral analysis of all
the sources within each system are tabulated below in each of the
individual galaxy sections. The point source spectra generally contained
low numbers of counts, and Gaussian statistics could not be assumed. A 
maximum likelihood criterion, as described above, was used.

\end{itemize} 

\subsection{HRI Observations and Data Reduction}
\label{sec_hri}

The three HRI datasets (NGC~520, NGC~2623 and NGC~7252) were analysed in
exactly the same way, again using the STARLINK {\em ASTERIX} X-ray
analysis system. In each case the data were first binned into a
0.4$^{\circ}\times0.4^{\circ}$ image with a 3\arcs\ resolution, thus
exploiting the HRI's superior spatial resolution. Background subtraction
of HRI data, although significantly easier than of PSPC data, is still
non-trivial, and two methods were utilised, which were compared both with
each other and with the mean HRI background rate ($\approx
1.1\times10^{-6}$\,ct s$^{-1}$ arcsec$^{-2}$; David \etal\ 1993).
Firstly, a polar profile of the data was formed about the centre of the
system, with radial bins of width 0.001$^{\circ}$. This was seen to level
out at large radius to a constant background level. The second method
involved an evaluation of the mean value, once all the bright features
had been removed from the data. The background values obtained from the
two methods agreed well in all three cases. These values were subtracted
from the raw data (over such a small area, the vignetting is essentially
flat, and can be ignored).

The resultant image was then exposure corrected, and searched for sources,
again using PSS. Again, a detection threshold of $4\sigma$ was assumed. 
Images, smoothed on many different spatial scales (from
3\arcs to 18\arcs), were formed  to search for extended low-surface brightness
features.

Count rates of the detected point sources within the three HRI datasets were converted into
fluxes (and then into luminosities), assuming 0.3\,keV and 3.0\,keV thermal bremsstrahlung
models. HRI conversion factors for both models from number of counts into fluxes (taking into
account the different exposure times and Galactic foreground $N_{H}$ values) are given in
Table~\ref{table_conv_fac}. 

\begin{table}
\begin{center}
\begin{tabular}{|l|c|c|}	\hline
System & BR & BR \\
       & 0.3\,keV & 3.0\,keV \\ \hline
NGC~520    & 6.47$\times10^{-4}$ & 3.50$\times10^{-4}$ \\ 
NGC~2623   & 4.54$\times10^{-4}$ & 2.31$\times10^{-4}$ \\
NGC~7252   & 7.11$\times10^{-4}$ & 4.99$\times10^{-4}$ \\
\hline
\end{tabular}
\caption[]{\small Counts-to-flux Conversion factors for the \Ros\ HRI detector for the
three individual HRI datasets and for two separate thermal bremsstrahlung models. The
values are corrected for Galactic absorption and are in units of $10^{-11}$\,erg
cm$^{-2}$ s$^{-1}$ cts$^{-1}$.}
\label{table_conv_fac}
\end{center}
\end{table}

Although the spectral response of the HRI is rather poor, it is thought
(Fraser 1992) that the ratio of the number of counts in channels 1$-$5 to
the number of counts in channels 6$-$11, used by Wilson \etal\ (1992),
does seem the most sensitive energy indicator that can be constructed.
The above analysis was also repeated for data within these two
energy bands.

\section{Results}
\label{sec_results}

The following eight subsections describe the results of the analyses of
\Ros\ PSPC and, where applicable, HRI data of the sample systems. Where
the analysis technique differs from that described in
Section~\ref{sec_obs_data} this is noted in the text.

Within each of Sections~\ref{sec_arp270} to \ref{sec_am1146-270}, a 
table is given, showing the main results of the analysis of each set of
\Ros\ PSPC data. The results of the spectral analyses of each of the
three types of spectra formed as described in Section~\ref{sec_obs_data} (an
integrated spectrum, a diffuse spectrum and individual source spectra), are
given. The number of counts, plus the statistical uncertainty, contained
within each spectrum are given, together with the results of the best spectral
fit to each spectrum. A quoted photon index indicates that the best fit is a 
power law model, whereas if instead, a temperature is given, the best fit is
either a Raymond \& Smith hot plasma model (if the metallicity is quoted), or
it is a bremsstrahlung model (if no metallicity is quoted). Where 
$\chi^{2}$-fitting has been used (in fitting most of the integrated spectra
plus a small number of other high-count rate cases), the values of the
$\chi^{2}$ statistic are given together with the number of degrees of freedom.
The (0.1$-$2.0\,keV) X-ray luminosities tabulated are those {\em escaping}
from the system in question, \ie before absorption in our own galaxy. Also
given in each case is the value of $L_{X}$(diffuse+sources), the sum of the
diffuse X-ray luminosity and each of the individual source luminosities.

Unless otherwise stated, we also present for each source a PSPC image
showing contours of background-subtracted, exposure-corrected X-ray emission
in the 0.1$-$2.3 keV band, superimposed on an optical image of the system.
These X-ray images have a resolution of 5\arcs\ and have been lightly smoothed
with a Gaussian of standard deviation 10\arcs\ to suppress noise. The contour
levels increase by factors of two from 7.2$\times10^{-4}$\,ct s$^{-1}$
arcmin$^{-2}$.

	\subsection{Arp~270}
	\label{sec_arp270}

Arp~270 (also VV~246; Vorontsov-Velyaminov, 1959) is composed of two
galaxies, thought to be of comparable masses (Davis \& Seaquist 1983),
NGC~3395, to the west, and the somewhat smaller, more irregular galaxy, NGC~3396,
to the east. The angular separation of their nuclei ($\approx90\arcs$)
corresponds to a separation of about 12\,kpc at the distance of 28\,Mpc
assumed within this paper. In the far-infrared, they appear as a strong,
though unresolved source (Soifer \etal\ 1987), and in the UV, they are
noted in the KISO catalogue (Takase \& Miyauchi-Isobe 1987) as an
interesting pair of UV-excess galaxies. Many radio continuum studies have
been made, the most noteworthy being that of Huang \etal\ (1994), who
find that the radio emission in each galaxy comes
predominantly from that part of each galaxy near its companion galaxy.
Whereas NGC~3396 shows two strong radio peaks and one weak one, its
partner has a much more complex structure, with six or more peaks in the
north-eastern part of the galaxy alone. A clear bridge between the two
galaxies is also seen in the radio, roughly coincident with the optical.

Arp~270 has been observed in X-rays previously to \Ros, with the \Ein\ IPC. The \Ein\ image 
(Fabbiano \etal\ 1992) shows very little structure, what there is appearing to be centred 
on NGC~3395. Suggestions of extensions are also visible to the north-west and south-west of the 
system. 

The \Ros\ image (Fig.~\ref{fig_arp270}) shows far more structure than
seen with \Ein. Emission is seen to come from both galaxies, and this
emission appears to be enshrouded in more diffuse emission. The X-ray feature
associated with the western galaxy (source~A; see
Table~\ref{results_arp270}) appears to be centrally positioned and lies
less than 2$\arcs$ from the position given in the Third Reference Catalog
of Bright Galaxies (RC3) (de Vaucouleurs
\etal\ 1991). Source~B however, does appear, like the radio emission discussed above, to emanate
from a position in NGC~3395 nearer to its companion galaxy (X-ray
source~B lies some 27$\arcs$ from the RC3 position of NGC~3396. As Huang
\etal\ (1994) point out, this is not expected from dynamical simulations.
Instead, the gas should be forced through the tidal interaction towards
the nuclear regions of each galaxy, and be compressed into
star-formation. This process however, as Huang \etal\ note,
will take some time, and the star-formation may well begin before the gas
reaches the nuclear regions. 
Non-nuclear bursts of star-formation can occur in the collision between two
gas-rich spiral galaxies, when a burst of massive star-formation is first
triggered in the overlapping/interpenetrating regions of the galaxies (Jog \&
Solomon 1992). This overlapping region will at first be the very outer discs
of the two galaxies. In the case of Arp~270 though, it appears that this very
initial phase may be nearly over, as gas is now seen close to the nuclear
regions, \ie the forcing of gas through the interaction towards the galactic
nuclei has begun. The fact that, in the case of NGC~3396, this hot gas is seen
only close to the nucleus, not yet within it, may indicate that this second
phase, the tidal forcing of gas towards the two nuclei, may have only just
begun.
Furthermore, there is quite a good correlation between the X-ray contours in
Fig.~\ref{fig_arp270} and the radio data of Huang \etal\ (1994) -- both the
tentative X-ray `bridge', seen apparently connecting the two galaxies, and the
radio bridge appear to connect the two systems along the southern edge of the
optical bridge.

\begin{figure} 
\vspace{80mm} 
\caption[]{\small
Contours of (0.1$-$2.3\,keV) X-ray emission shown superimposed on an optical image of
Arp~270. Contour levels increase by factors of two from $7.2\times10^{-4}$\,ct
s$^{-1}$ arcmin$^{-2}$.)}
\label{fig_arp270}
\end{figure} 

Both source spectra are best fit with low-temperature, absorbed plasma models, suggestive of
the emission from both regions being due to hot gas, rather than to evolved stellar
components. The fact that no evidence was found in the radio for an active galactic nucleus in
either galaxy, and that the radio spectra of the galaxies are relatively flat (at low
frequencies) (Huang \etal\ 1994), indicative of a good deal of supernova activity, adds
credence to the hot gas idea.

The two other detected sources in the field, the strong point-like source
to the north of NGC~3396, and the weaker source within the X-ray `tail'
to the south-west of NGC~3396, appear to explain the suggested extensions
tentatively seen in the \Ein\ image. The northern source is unlikely to
have anything to do with the system, and is most probably a background
quasar, its best spectral fit being a power-law fit with a photon index,
$\alpha=2.3\pm0.5$, consistent with the mean spectrum of quasars in the
\Ros\ band (2.2$\pm$0.2; Branduardi-Raymont
\etal\ 1994; Roche \etal\ 1995). The X-ray feature 
to the south-west of NGC~3395 however, may well be
associated with the system. The radio observations of Huang \etal\ (1994)
do show that some emission has diffused beyond the optical edge of the
galaxy within this region, perhaps indicative of previous supernova
outbursts. This would help to explain the existence of hot X-ray emitting
gas also. The number of counts within this region make
it difficult to constrain its spectral properties.

The `diffuse' spectrum of Arp 270 as a whole appears severely
contaminated, both by unresolved point sources, and by `stumps' of source
emission left in the source-subtraction procedure (see
Section~\ref{sec_pspc}). This is reflected in the rather wide confidence
errors quoted on the best fit values in Table~\ref{results_arp270}
(remember, no goodness of fit can be quoted as a maximum likelihood
technique had to be used because of the small number of counts). As
discussed in Section~\ref{sec_pspc}, we attempted to fit a two-component
model to the diffuse spectrum, comprising a Raymond \& Smith hot plasma
(representing the hot gas), and a highly absorbed hot (10\,keV) component
(representing the source contamination component). This two-component fit
suggests, as shown in Table~\ref{results_arp270}, that approximately
two-thirds of the `diffuse' emission may be truly diffuse gas at a very
low ($<0.2$\,keV) temperature, while the remaining third of the emission
can be accounted for in terms of highly absorbed, hard (10\,keV) sources.
An F-test shows that the improvement in fit quality to the `diffuse'
spectrum from including the second component is significant at over 99\%
confidence.

\begin{table*}
\begin{center}
\begin{tabular}{cccccccccc}	\hline
\multicolumn{3}{c}{Source Spectrum Analysed} & Net  & $\log{L_{X}}$ &\multicolumn{5}{c}{Best
	Spectral Fit Results}  	\\ \cline{6-10}
    &R.A. & Dec &Counts& erg s$^{-1}$ & column & photon & temp. & Z
	& $\chi^{2}$ (n.d.o.f) \\ 
    &(2000.0)&(2000.0)&      &(0.1$-$2.0\,keV)& $10^{20}$\,cm$^{-2}$ &
	index & keV & (solar) & \\ \hline 
\multicolumn{3}{c}{Integrated} & 396$\pm$37.4 & 40.48 & 7.76$^{+2.94}_{-2.03}$ &
3.11$^{+0.44}_{-0.43}$ & & & 20.16 (19) \\ \\
\multicolumn{3}{c}{`Diffuse' 1-component fit} & & 40.04 & 5.10$^{+5.46}_{-3.18}$ & 
 & 2.56$^{+9.51}_{-0.30}$ & 0.00$^{+0.38}_{-0.00}$ & \\ 
\multicolumn{3}{c}{`Diffuse' 2-component fit} & 168$\pm$34.2 & (RS) 40.01 & 2.01$^{+72.1}_{-0.48}$ & 
 & 0.17$^{+0.11}_{-0.11}$ & 1.30$^{+5.42}_{-0.44}$ & \\ 
\multicolumn{3}{c}{                         } & & (Br) 39.75 & 150$^{+93.4}_{-71.7}$ & 
 & 10 (F) & & \\ \\
Source~A & 10 49 49.58 & +32 58 51.6 & 112$\pm$13 & 39.92 &
	6.49$^{+3.02}_{-1.78}$ & & 0.57$^{+0.21}_{-0.13}$ & &  \\
Source~B & 10 49 54.53 & +32 59 16.9 & 95.3$\pm$12 & 39.81 &
	4.91$^{+2.30}_{-1.60}$ & & 0.63$^{+0.30}_{-0.17}$ & &  \\ \hline
\multicolumn{4}{c}{Best `Diffuse' + sources} & 40.49 &
& & & & \\ 
\hline
\end{tabular}
\caption[]{\small The results of the PSPC spectral analysis of Arp~270. 
Tabulated luminosities are those escaping from the system. Only the best model fit is shown.
Errors on the spectral fit parameters are 1$\sigma$ for one interesting parameter.}
\label{results_arp270}
\end{center}
\end{table*}

	\subsection{Arp~242}
	\label{sec_arp242}

One of the original systems presented by Toomre \& Toomre (1972) as a classic example of a pair
of galaxies undergoing tidal interaction is Arp~242 (the {\em Mice}). Also known as NGC~4676,
this system lies second in the proposed evolutionary sequences of both Toomre (1977) and Hibbard
\& van Gorkom (1996), as well as in the present work. Recent kinematical work (e.g. Mihos,
Bothun \& Richstone 1993) has strengthened further the tidal interpretation of the {\em
Mice}'s structure, and it is now generally agreed that both galaxies, in the throes of a
prograde encounter, have their northern edges moving away from us, such that NGC~4676a's tail
is on the very furthest side, swinging away from us, and NGC~4676b is rotating clockwise, its
northeastern portion, the closest to us (Hibbard \& van Gorkom 1996). Hibbard \& van Gorkom
(1996) find these kinematics in general agreement with their \hi\ velocity data. What
discrepancies they do see allow them to derive a mass ratio of 2:1 for NGC~4676b:NGC~4676a.

Both the northern galaxy (NGC~4676a) and the southern galaxy (NGC~4676b) appear to have shapes
and colours consistent with those of early-type spirals, though the disc regions are strongly
distorted or absent. Both tails, although bluer than the galaxies' central colours, are in
agreement with the colours of outer-disc regions (Schombert, Wallin \& Struck-Marcell 1990).
They account for 16\% of the total $H{\alpha}$ emission, are quite luminous, containing
one-third of the total R-band luminosity of the system, and have a high atomic gas content
(Hibbard \& van Gorkom 1996).

The northern galaxy appears to exhibit a 6.6$h^{-1}$\,kpc plume of $H{\alpha}$ along its minor
axis, and the southern galaxy possesses an ionized gas bar, as produced in Barnes \& Hernquist's
(1991, 1996) merger simulations, offset with respect to the stellar bar (Hibbard \& van
Gorkom 1996). Angular momentum transfer between the two bars is able to force large amounts of
gas towards the galactic centre.

Arp 242 has never been observed in X-rays previously. The X-ray image
(Fig~\ref{fig_arp242}) shows rather an amorphous X-ray structure. The one
source detected (labelled `A') lies further than the \Ros\ rms attitude uncertainty
($\approx6\arcs$; Hasinger \etal\ 1992) from the nuclei of either galaxy,
lying approximately 40$\arcs$ south of NGC~4676a and 20$\arcs$ west of
NGC~4676b. It may be, that the source is actually associated with the
contact region between the two galaxies, and is either collisionally
heated gas or possibly due to enhanced star formation taking place in the
dense molecular clouds in the contact region. Unfortunately, due to lack
of counts and limitations on resolution, very little can be said about
the spectral properties of the source, or its true position.

Leaving aside this source for the moment, the remainder of the emission does seem, in general,
to follow the optical `heads' of the galaxies, running from the south-east to the north-west.
The little `spur' to the south-west, though interesting, is probably spurious, or if not,
unlikely to be associated with the system. The feature to the north-west however, may well be
associated with the NGC4676a $H{\alpha}$ plume (Hibbard \& van Gorkom 1996) extending in the
same direction. The correlation of $H{\alpha}$ and X-ray features extending along the minor
axis of galaxies, is often a good indicator of a starburst-driven galactic wind taking place,
both the $H{\alpha}$ and the X-ray emission being due to clouds (of different densities)
being shocked by the hot, fast wind from the galactic nucleus (Heckman, Lehnert \& Armus 1993). 
While the fitted temperature to the diffuse emission spectrum is in very good agreement with
fitted temperatures of more nearby, known, starburst winds (Heckman 1993; Read \etal 1997),
the small number of counts, make it impossible to be confident of this result.

\begin{figure} 
\vspace{80mm} 
\caption[]{\small
Contours of (0.1$-$2.3\,keV) X-ray emission shown superimposed on an optical image of
Arp~242. Contour levels increase by factors of two from $7.2\times10^{-4}$\,ct
s$^{-1}$ arcmin$^{-2}$.)}
\label{fig_arp242}
\end{figure}

\begin{table*}
\begin{center}
\begin{tabular}{cccccccccc}	\hline
\multicolumn{3}{c}{Source Spectrum Analysed} & Net  & $\log{L_{X}}$ &\multicolumn{5}{c}{Best
	Spectral Fit Results}  	\\ \cline{6-10}
    &R.A. & Dec &Counts& erg s$^{-1}$ & column & photon & temp. & Z
	& $\chi^{2}$ (n.d.o.f) \\ 
    &(2000.0)&(2000.0)&      &(0.1$-$2.0\,keV)& $10^{20}$\,cm$^{-2}$ &
	index & keV & (solar) & \\ \hline 
\multicolumn{3}{c}{Integrated} & 53.8$\pm$19.0 & 40.76 &
	4.37$^{+5.53}_{-4.37}$ & &0.10$^{+0.13}_{-0.10}$  & 0.05$^{+118}_{-0.05}$
	& 41.78 (18) \\ \\
\multicolumn{3}{c}{Diffuse}    & 31.2$\pm$18.4 &          40.66 &
	0.00$^{+2.63}_{-0.00}$ & 
	 & 0.35$^{+0.39}_{-0.19}$ & 0.04$^{+0.45}_{-0.04}$ & \\ \\
Source~A & 12 46 09.96 & +30 43 20.8 & 17.5$\pm$5.5 & 40.22 &
	66.2$^{+121}_{-63.3}$ & & 0.24$^{+0.58}_{-0.15}$ & &  \\ \hline
\multicolumn{4}{c}{Diffuse + sources} & 40.80 &
& & & & \\ 
\hline
\end{tabular}
\caption[]{\small The results of the PSPC spectral analysis of Arp~242.
Tabulated luminosities are those escaping from the system. Only the best model fit is shown.
Errors on the spectral fit parameters are 1$\sigma$ for one interesting parameter.}
\label{results_arp242}
\end{center}
\end{table*}

	\subsection{NGC~4038/9}
	\label{sec_ngc4038}

The {\em Antennae}, NGC~4038/9 (also Arp~244) is perhaps the classic example of a pair of
galaxies in gravitational interaction, with spectacular tails spanning nearly 150\,kpc. It has
been the study of many dynamical models (Toomre \& Toomre 1972; Barnes 1988), and the basic
validity of these models has been confirmed by studies in neutral (van der Hulst 1979) and
ionised (Amram \etal\ 1992) hydrogen gas. $H{\alpha}$ emission knots (Rubin, Ford \& Dodorico
1971), coincident with powerful radio features (Hummel \& Van der Hulst 1986; Wolstencroft 1988)
are found throughout the central parts of both galaxies, and Van der Hulst (1979) showed that
about 70\% of the \hi\ in the system lies in the optical tails. Both infrared (Bushhouse \&
Werner 1990; Vigroux \etal\ 1996) and CO (Stanford \etal\ 1990) observations indicate that
both the nuclei, and the contact region between the two galaxies, are likely to be sites of
very active star formation.

The results of the \Ein\ observation of the {\em Antennae} (Fabbiano \&
Trinchieri 1983) were rather inconclusive, though the emission was seen
to be extended, and contained a soft X-ray contribution. Harder emission
and a possible hard point source at the contact region were also seen.
The \Ros\ PSPC observation sheds far more light on to the matter, and full
details of the observation, the analysis and the results, are given in Read
\etal\ (1995), though a summary is presented below.

The \Ros\ PSPC data were processed initially in a manner very similar to the
method presented here, though further, more sophisticated techniques
were used in the latter stages 
of the analysis (see Read \etal\ 1995). Using a maximum entropy reconstruction
technique (Gull 1989; Skilling 1989), a number of discrete components to the
X-ray emission were identified (A$-$G; see Table~\ref{results_ngc4038} and
Fig.~\ref{fig_n4038}). Many of these have counterparts at other wavelengths; B
and D are associated with the northern (NGC~4038) and the southern (NGC~4039)
nuclei respectively. Their properties, the facts that they are best fitted with
low-temperature and relatively low-column plasma models, that there are
offsets between the X-ray positions and the radio positions, and that it is
known that there is a great deal of supernova activity taking place at these
sites, suggest that they are bubbles of hot extranuclear gas, rather than the
central starbursts themselves. Features A, C, and possibly E have massive
$H{\alpha}$ knots and radio features as counterparts, and are likely to be
giant \hii\ regions.

Nearly half of
the {\em Antennae}'s X-ray flux appears diffuse, and a two-component
spectral fit (see Table~\ref{results_ngc4038}), indicates that almost all
of it is genuine, hot ($4\times10^{6}$\,K), low-metallicity gas (the rest
being due to unresolved sources), with a total mass perhaps exceeding
$10^{9}$\,$M_{\odot}$. The bulk of this emission appears to envelope the
entire optical system, except for the tidal tails. An enhancement
(accompanied by a rise in temperature to over $10^{7}$\,K) is seen in
this diffuse emission at the contact point between the two galaxies,
where it is believed the two discs are colliding. This may be due to gas
heated by the collision, or to triggered star-formation in the vicinity.
Of great interest are the two elongated structures extending to the
northwest and to the southwest, extending from the galactic discs to
radii of $\sim30$\,kpc, and culminating in symmetrically-positioned,
apparently point-like, sources (P and Q). These streamers most likely indicate
the existence of galactic winds, their normal bipolar structure possibly
having been disturbed by the rapid dynamical evolution of the system
(see Section~\ref{sec_disc_diff} for further discussion of this).
Containing $\approx10$\% of the total hot gas
present, these streamers could represent a significant loss in mass and
energy from the system. 

The point sources at the ends of the two arms are a real puzzle. The sources
could be related to the arms if they were massive ($\sim10^{8} M_{\odot}$)
objects, ejected from the galactic nuclei (see Saslaw, Valtonen \& Aarseth
1974), the ejection process giving rise to the juxtaposed `wakes' of hot
gas. However there are various difficulties with this idea (Read \etal\
1995). Alternatively they could be foreground or background objects, for
example quasars or stars, completely unrelated to the diffuse arms. This latter
possibility is supported by recent optical observations of source~P, the optical feature
coincident with the source at the end of the northern streamer, which indicate that it
is a distant ($z=0.155$) galaxy with an active nucleus (L.R.Jones, private
communication).

The {\em Antennae} has also been observed with the \Ros\ HRI (Fabbiano,
Schweizer \& Mackie 1996), and the results are generally consistent with
the PSPC results. A complex X-ray structure is seen involving filamentary
regions, closely following the $H{\alpha}$ distribution, emission peaks
coincident with \hii\ regions, super-Eddington sources, and prominent
nuclear sources. There is also some evidence of nuclear outflows and
superbubbles.

\begin{figure} 
\vspace{80mm} 
\caption[]{\small
Contours of (0.1$-$2.3\,keV) X-ray emission shown superimposed on an optical image of
Arp~4038. Contour levels increase by factors of two from $7.2\times10^{-4}$\,ct
s$^{-1}$ arcmin$^{-2}$.)}
\label{fig_n4038}
\end{figure}

\begin{table*}
\begin{center}
\begin{tabular}{cccccccccc}	\hline
\multicolumn{3}{c}{Source Spectrum Analysed} & Net  & $\log{L_{X}}$ &\multicolumn{5}{c}{Best
	Spectral Fit Results}  	\\ \cline{6-10}
    &R.A. & Dec &Counts& erg s$^{-1}$ & column & photon & temp. & Z
	& $\chi^{2}$ (n.d.o.f) \\ 
    &(2000.0)&(2000.0)&      &(0.1$-$2.0\,keV)& $10^{20}$\,cm$^{-2}$ &
	index & keV & (solar) & \\ \hline 
\multicolumn{3}{c}{Integrated} & 2804$\pm$61 & 40.98 & 5.50$^{+0.50}_{-0.40}$ &
 & 0.69$^{+0.03}_{-0.03}$ & 0.03$^{+0.01}_{-0.01}$ & 23.7 (14) \\ \\
\multicolumn{3}{c}{`Diffuse' 1-component fit} & &  40.63 & 3.30$^{+0.87}_{-0.77}$ & 
 & 0.49$^{+0.07}_{-0.06}$ & 0.04$^{+0.02}_{-0.02}$ & 21.56 (14) \\ 
\multicolumn{3}{c}{`Diffuse' 2-component fit} & 521$\pm$32 & (RS) 40.61 & 3.40$^{+0.58}_{-0.00}$ & 
 & 0.36$^{+0.09}_{-0.05}$ & 0.07$^{+0.05}_{-0.03}$ & 17.80 (12) \\ 
\multicolumn{3}{c}{                         } & & (Br) 39.51 & 87.5$^{+215}_{-87.5}$ & 
 & 10 (F) & & \\ \\
Source~A & 12 01 55.4 & -18 52 09 & 475$\pm$23 & 40.28 & 
	8.1$^{+1.4}_{-1.1}$ & & 0.72$^{+0.13}_{-0.10}$ & 0.00$^{+0.01}_{-0.00}$  &  \\
Source~B & 12 01 52.2 & -18 52 09 & 129$\pm$12 & 39.99 & 
	5.1$^{+1.6}_{-1.3}$ & & 0.79$^{+0.13}_{-0.13}$ & 0.06$^{+0.09}_{-0.04}$  &  \\
Source~C & 12 01 50.8 & -18 52 20 & 124$\pm$12 & 39.91 & 
	17.4$^{+14.4}_{-7.4}$ & & 0.68$^{+0.26}_{-0.26}$ & 0.02$^{+0.06}_{-0.02}$  &  \\
Source~D & 12 01 53.3 & -18 53 06 & 520$\pm$24 & 40.16 & 
	13.7$^{+5.8}_{-3.4}$ & & 0.70$^{+0.06}_{-0.10}$ & 0.10$^{+0.07}_{-0.05}$  &  \\
Source~E & 12 01 51.2 & -18 53 45 & 119$\pm$12 & 39.70 & 
	10.5$^{+15.8}_{-4.0}$ & 2.2$^{+1.1}_{-0.5}$ & & &  \\
Source~F & 12 01 47.9 & -18 54 10 & 53$\pm$9   & 38.98 & 
	1.9$^{+5.6}_{-1.9}$ & & 0.45$^{+0.71}_{-0.17}$ & 0.02$^{+0.64}_{-0.02}$  &  \\
Source~G & 12 01 53.5 & -18 51 23 & 117$\pm$13 & 39.42 & 
	11.5$^{+20.4}_{-4.5}$ & & 0.38$^{+0.17}_{-0.15}$ & 0.00$^{+0.02}_{-0.00}$  &  \\ \hline
\multicolumn{4}{c}{Best `Diffuse' + sources } & 41.03 &
& & & & \\ 
\hline
\end{tabular}
\caption[]{\small The results of the PSPC spectral analysis of NGC~4038/9.
Tabulated luminosities are those escaping from the system. Only the best model fit is shown.
Errors on the spectral fit parameters are 1$\sigma$ for one interesting parameter.}
\label{results_ngc4038}
\end{center}
\end{table*}

	\subsection{NGC~520}
	\label{sec_ngc520}

The nature of the peculiar system NGC~520 was once a puzzle; is it one
disturbed galaxy or two interacting galaxies? Recent studies have
clarified the interacting galaxies interpretation. Stanford and Balcells
(1990) have detected two galactic nuclei, just visible in the optical
but more clearly in their K-band image. The less massive component (by perhaps
more than an order of magnitude) is the northwestern knot, which is
optically brighter than the main component. The main component is
optically weak because it is seen edge-on, and the light from its central
region is absorbed by interstellar dust in its disc (visible as a dark
lane in Figure~1 of Bernl\"{o}hr 1993). Furthermore, Two hypotheses,
either that the nearby dwarf galaxy UGC~957 might be primarily
responsible for the disturbed morphology of a single galaxy in NGC~520,
or that two interacting disc systems formed NGC~520, were tested with
numerical simulations (Stanford and Balcells 1991). The simulations
indicate that NGC~520 contains two interacting discs which collided
$\sim3\times10^{8}$ years ago and UGC~957 was only involved in the
producing of the northern half of a tidal tail.

Tovmasyan and Sramek (1976) found that the compact radio source in NGC~520 is situated in the
dark lane between the two visible parts of the system. Condon \etal\ (1982) later resolved this
into a $6''$ east-west extension, consistent with an edge-on disc, lying $3^{\circ}$ off
east-west. Millimeter-wave interferometer maps of the 2.6$\mu m$ CO emission (Sanders \etal\
1988a) show a strong peak at the position of this radio source; approximately $1.9\times10^{9}
M_{\odot}$ of molecular gas is concentrated in a region approximately 0.8\,kpc in size.

Much of the extranuclear regions of both galaxies within NGC~520 experienced a period of
enhanced star formation $\sim3\times10^{8}$ years ago. The main sequence remnants of this burst
are the A stars whose features are evident in the optical spectra (Stanford 1991). The burst
within the less massive (the north-western) nucleus occurred slightly later than in the
extranuclear regions, but the star formation rate has returned to a nominal level. The more
massive, optically hidden nucleus (to the south-east) produces stars at a rate of $\sim0.7
M_{\odot}$\,yr$^{-1}$ and is the current dominant source of star formation in this system. The
star formation rate within this region is $\sim35$ times higher than for an isolated disc
galaxy. This region dominates the mid-infrared flux of the system, and probably produces most
of the far-infrared flux seen in NGC~520. It is still unclear, though, whether the massive star
component is sufficient to power the activity at the very centre, even though multi-wavelength
observations argue against the presence of a nonstellar compact power source within the
south-eastern nucleus (Stanford 1991).

The NGC~520 PSPC data were
processed exactly as described in section~\ref{sec_obs_data}.
Figure~\ref{fig_ngc520} shows contours of (0.1$-$2.3) X-ray emission, superimposed on an
optical image. Table~\ref{results_ngc520} shows the results of the spectral fitting to the PSPC
data as described in Section~\ref{sec_obs_data}. The one source detected in the vicinity of
NGC~520 at $\alpha=01h24m34.76s$, $\delta=+03d47m39.7s$, lies within $\sim5''$ of the radio
source resolved by Condon \etal\ (1982), once the residual error of the \Ros\ attitude solution
($\approx6$\arcs; Hasinger \etal\ 1992) is taken into account.

\begin{table*}
\begin{center}
\begin{tabular}{cccccccccc}	\hline
\multicolumn{3}{c}{Source Spectrum Analysed} & Net  & $\log{L_{X}}$ &\multicolumn{5}{c}{Best
	Spectral Fit Results}  	\\ \cline{6-10}
    &R.A. & Dec &Counts& erg s$^{-1}$ & column & photon & temp. & Z
	& $\chi^{2}$ (n.d.o.f) \\ 
    &(2000.0)&(2000.0)&      &(0.1$-$2.0\,keV)& $10^{20}$\,cm$^{-2}$ &
	index & keV & (solar) & \\ \hline 
\multicolumn{3}{c}{Integrated} & 100.2$\pm$10.5 & 39.94 & 53.7$^{+47.7}_{-30.5}$ &
4.52$^{+2.83}_{-1.95}$ & & & 22.53 (19) \\ \\
\multicolumn{3}{c}{Diffuse}    & 19.4$\pm$10.1 & 39.12 &
0.00$^{+83.1}_{-0.00}$ & 
 & 0.43$^{+0.21}_{-0.10}$ & 10.0$^{+90.0}_{-7.13}$ & \\ \\
Source~A & 01 24 34.76 & +03 47 39.7 & 84.7$\pm$9.8 & 39.92 &
	9.27$^{+11.1}_{-3.86}$ & 1.90$^{+0.80}_{-0.56}$ & & &  \\ \hline
\multicolumn{4}{c}{Diffuse + sources } & 39.99 &
& & & & \\ 
\hline
\end{tabular}
\caption[]{\small The results of the PSPC spectral analysis of NGC~520. 
Tabulated luminosities are those escaping from the system. Only the best model fit is shown.
Errors on the spectral fit parameters are 1$\sigma$ for one interesting parameter.}
\label{results_ngc520}
\end{center}
\end{table*}

\begin{figure} 
\vspace{80mm} 
\caption[]{\small
Contours of 0.1$-$2.3\,keV X-ray emission from NGC~520 overlayed on an
optical image. The contour levels increase by factors of two from
$7.2\times10^{-4}$\,ct s$^{-1}$ arcmin$^{-2}$.} 
\label{fig_ngc520}
\end{figure} 

It appears that NGC~520 is a very compact X-ray source with little in the way of diffuse
emission. The \Ein\ image (Fabbiano \etal\ 1992) appears to show the same thing though to a
much coarser resolution. This compactness can be seen more clearly in
figure~\ref{fig_ngc520_radial}, a comparison of the radial profile of the emission from NGC~520
(crosses) with the \Ros\ PSPC PSF (line), where very little emission is seen beyond a radius of
0.6\arcm\ (the 95\% enclosed energy radius of the PSPC PSF). The X-ray emission from NGC~520 is
almost consistent with point source emission, and what diffuse emission exists, only makes up a
very small fraction of the total.

\begin{figure} 
\vspace{70mm} 
\caption[NGC~520: a comparison of the radial X-ray emission with the \Ros\ 
PSPC PSF.]{\small A comparison of the radial X-ray emission from NGC~520 
(crosses) with the \Ros\ PSPC PSF (line).}
\label{fig_ngc520_radial}
\end{figure} 

The diffuse emission that could be extracted (using the method described in
section~\ref{sec_obs_data}) amounts to only about 20 counts, and little can be said about its
spectral properties, except that it appears soft, perhaps suggestive of a very low-level, cool
corona surrounding the system. This apparent lack of diffuse emission in NGC~520 is discussed
in detail later in Section~\ref{sec_evolution}.

The one point source detected, coincident with both the compact radio source resolved by Condon
\etal\ (1982), and with the more massive of the two nuclei, as visible in the K-band image of
Stanford \& Balcells (1990), appears to dominate the X-ray emission. The power-law fit and the
lack of diffuse emission perhaps point to an AGN as the primary nuclear power source, but
NGC~520 is, as shall be seen in Section~\ref{sec_evolution}, far too X-ray dim.
No source is detected at the position of the north-western nucleus, though an elongation of
the X-ray contours in this direction is tentatively visible in figure~\ref{fig_ngc520}.

In contrast to the PSPC image, the HRI image (Fig.~\ref{fig_ngc520_hri}) shows a great
deal of detail. Table~\ref{results_ngc520_hri} gives the positions, net counts and X-ray
luminosities (assuming 0.3\,keV (and 3.0\,keV) bremsstrahlung models) of the detected HRI
sources. Three sources are detected, the most northerly, H1, being centred less than
5$\arcs$ (\ie less than the residual error in the \Ros\ attitude solution) east of the
secondary (optically brighter) nucleus. This source, undoubtedly associated with the
northern nucleus, appears to be the hardest of the sources, as it is the only source
detected in the hard band (channels 6$-$11; see Section~\ref{sec_hri}). The suggested
extension to the east is interesting, though its existence is tentative. Sources H2 and
H3 both appear to be much softer than source H1, and they follow the bright band of
optical emission down the north-eastern side of the system. The primary, optically
fainter, southern nucleus actually lies slightly south of H2, and west of H3, where 
some X-ray enhancement is visible.

\begin{figure} 
\vspace{80mm} 
\caption[]{\small
Contours of \Ros\ HRI X-ray emission from NGC~520 overlayed on an
optical image. The contour levels increase by factors of two from
$2.2\times10^{-3}$\,ct s$^{-1}$ arcmin$^{-2}$.} 
\label{fig_ngc520_hri}
\end{figure} 

\begin{table}
\begin{center}
\begin{tabular}{ccccc}	\hline
       & R.A. & Dec. & Net & $\log{L_{X}}$ \\
       & (2000.0) & (2000.0) & counts & (erg s$^{-1}$) \\ \hline
H1     & 01 24 33.51 & +03 48 03.5 & 9.60$\pm$4.5 & 39.77 (39.50) \\
H2     & 01 24 34.71 & +03 47 41.2 & 10.3$\pm$4.9 & 39.80 (39.53) \\
H3     & 01 24 35.71 & +03 47 31.5 & 9.59$\pm$4.8 & 39.77 (39.50)\\
\hline
\end{tabular}
\caption[]{\small Sources within NGC~520 detected with the HRI. Luminosities are calculated 
as described in Section~\ref{sec_hri}, using a 0.3\,keV (and a 3.0\,keV) thermal 
bremsstrahlung model, corrected for Galactic absorption.}
\label{results_ngc520_hri}
\end{center}
\end{table}

	\subsection{Arp~220}
	\label{sec_arp220}

Arp~220 (also IC~4553, UGC~9913) is the prototypical ultraluminous galaxy, and is the most
luminous object in the local universe. Being the closest and brightest of these IRAS-discovered
(Sanders \etal\ 1988b) systems, it has been extensively studied across the whole range of the
electromagnetic spectrum. The origin of Arp~220's intense power however, remains still somewhat
of a mystery. On the one hand, the huge quantities of dense, warm molecular gas (Scoville \&
Soifer 1991), together with the strong stellar CO absorption lines observable in the IR (Armus
\etal\ 1995), point to a massive burst of star-formation as the primary energy source. On the
other, there appears to be a lot of indirect evidence for the presence of
a quasar obscured by a layer of dense gas and dust. Lonsdale \etal\
(1994), for instance, report a OH megamaser in Arp~220 with an OH line
peak originating in a structure $\ltsim$1\,pc across, and they conclude
that most of the emission from Arp~220 arises from a very small region,
perhaps on a scale of less than 10\,pc, hence the quasar nucleus
hypothesis. Prestwich, Joseph \& Wright (1994) suggest that, while the IR
spectrum of Arp~220 is well fit by a starburst model (Rowan-Robinson \&
Crawford 1989), its relative deficit of ionizing photons may indicate
that it is an evolved starburst, now dominated by an active nucleus. They
also point out though, that extinction plus dust absorption of the
Ly${\alpha}$ photons could result in a much decreased recombination line
flux. A third hypothesis for the production of such large amounts of
power, as discussed below, involves the idea of a galaxy-galaxy
collision. There is much evidence for Arp~220 being a merger of two
galaxies. Majewski \etal\ 1993, for instance, have observed the very
close, double nucleus in the near-IR. The optical morphology
is very similar to that seen in dynamical simulations of late-stage mergers of
two disc galaxies (\eg Mihos \& Hernquist 1994). Furthermore, Wright
\etal\ (1990), having obtained K-band images of several merging systems,
discovered that the radial profile of Arp~220 appeared to follow an
$r^{1/4}$ law, typical of elliptical-like, and therefore possibly
merger-remnant, systems. Within this merger scenario, it has been
suggested (Harwitt \etal\ 1987), that the FIR emission and the soft
X-ray/EUV radiation could be produced by dust heated by the collision,
and by the cloud-cloud collisions that occur when two-gas-rich systems
encounter one another.

Arp~220 has been observed in X-rays prior to \Ros, with the \Ein\ IPC (Eales \& Arnaud 1988). 
An X-ray sources $\sim1\arcm$ from the position of Arp~220 was discovered, and was assumed to be
associated with the system. Because of the size of the source (thought to be about twice the
optical size) and the softness of the spectrum, the authors concluded that the bulk of the
emission could not be coming from any obscured active nucleus at the galaxy's centre. 

A thorough analysis of both the \Ros\ PSPC and HRI data from Arp~220 has
already been published (Heckman \etal\ 1996), and a discussion of our
results in conjunction with theirs, is presented here. The \Ros\ image
(Fig.\ref{fig_arp220}) shows quite clearly why the \Ein\ image was so
confusing. Two distinct regions of X-ray emission are seen around
Arp~220. One is centred on the system itself, and appears elongated in an
approximately east-west direction. The other feature is a large
($\sim3\arcm$) amorphous structure, and appears to stretch out from
Arp~220 to the south and south-west. Two sources are detected. Source~A
lies within the large southern feature and source~B lies at the centre of
Arp~220.

The northern feature, the approximately east-west structure centred on
source~B, at the centre of Arp~220, is large, $\approx40$\,kpc long at
the distance of 76\,Mpc assumed here. Interestingly, Heckman \etal's
(1996) HRI image also shows an elongation ($\approx11$\,kpc long), though
in a more southeast-northwest orientation, misaligned by about
30$^{\circ}$ with the PSPC feature. The HRI image also shows, at the very
centre of the emission, what appears to be a pair of pointlike sources of
equal brightness, separated by $\approx5\arcs$, lying in a
southeast-northwest orientation, similar to what is seen in optical
R-band and H${\alpha}$ images. It seems very likely that this
double structure is related 
to the prominent dust lane running through the centre of the galaxy.
The X-ray structure is probably an elongated feature which appears as
a double source because of the dust lane running perpendicularly through
its centre.

The northern east-west elongated structure appears very similar to the
H${\alpha}$ emission-line nebula visible around Arp~220 (Heckman, Armus
\& Miley 1987; 1990), which, as shown by Heckman \etal\ (1990), appears
to be a pair of expanding bubble-like structures. As noted by Heckman
\etal\ (1996), both the X-ray and the optical features show a central
maximum, surrounded by a relatively bright, 0.5\arcm\ inner region,
elongated in the southeast-northwest direction. Outside this, a larger
structure is visible in both the X-ray and the optical, elongated more in
the east-west direction. Heckman \etal\ (1996) conclude that both the
X-ray and the H${\alpha}$ features are the same size (to within 10\%).
They also suggest that both the X-ray and the optical emission arise as
the result of a bipolar wind, driven out from the nucleus over a
timescale of a few times $10^{7}$\,yr by an ultraluminous starburst
(though it is also possible that the wind could be driven by a hidden
QSO). Heckman \etal\ (1996) also suggest that the $\sim30^{\circ}$
misalignment of the outer structure may reflect a change in orientation of the
system, as the encounter has progressed (see
Section~\ref{sec_disc_diff_mid}).

In the previous analyses, the X-ray features seen have not been much
larger than the optical sizes of the systems in question (except perhaps
in the case of the {\em Antennae}, where a more rigorous analysis has
been performed; see Section~\ref{sec_ngc4038}). In Arp~220 however, the
X-ray emission extends well beyond the optical confines of the system,
and so the spectral extraction procedure has been performed in a
different way to that described in Section~\ref{sec_obs_data}. Firstly, a
`northern' spectrum was formed, in an identical manner to that described
in Section~\ref{sec_obs_data}, from a circular area slightly greater than
the optical diameter (actually centred on source~B and extending out to a
radius of 65\arcs). Source spectra for sources~A and B have been formed
exactly as described in Section~\ref{sec_obs_data}. The `diffuse'
spectrum however, has been extracted from a much larger area, chosen to
include the southern emission -- a circle of 3\arcm\ radius, centred
halfway between sources~A and B. The emission from sources~A and B has
been removed, as described in Section~\ref{sec_obs_data}.

The `northern' spectrum and its resultant fits (see
Table~\ref{results_arp220}), therefore refer to the whole of the northern
feature, the east-west feature surrounding (and including) source~B.
Heckman \etal\ (1996), in their fitting of the spectrum obtained from
this area, froze the absorbing hydrogen column at two values; the column
out of our own galaxy (3.7$\times10^{20}$\,cm$^{-2}$), and the maximum
column allowable by the \hi\ data of Baan
\etal\ (1987) (5$\times10^{21}$\,cm$^{-2}$). Neither fit can be excluded.
The best fit obtained in our present work (though relatively bad) is
consistent with the high-column fit of Heckman \etal\ (1996).
Consequently, our derived $L_{X}$(0.1$-$2.0\,keV) of
$1.5\times10^{41}$\,erg s$^{-1}$ is towards the higher side of the
Heckman \etal\ (1996) range. Little can be said about the different
spectral components involved in the northern feature except that the
central peak (source~B), is best fitted with a very high-column plasma,
and may be spectrally harder than the total `northern' emission,
indicating that the outer, non-source~B emission, - the `bubble'
emission, is likely to be spectrally, quite soft, as is seen in the halos
of many starburst and merging systems (\eg\ Heckman 1993; Read \etal\
1997).

The southern feature is very large, about 3\arcm\ (65\,kpc) in diameter, bright, accounting for
about half of the total counts, and appears to be spectrally, quite soft - the best
fit to the `diffuse' spectrum (\ie the entire emission minus the emission from sources~A and
B) (Table~\ref{results_arp220}) being consistent with a low temperature (6 million K) plasma,
absorbed merely by the column within our own galaxy. 

Heckman \etal\ (1996) suggest that this southern feature is associated
with a background group/poor cluster of galaxies, though
they cannot exclude the possibility that it is in fact, associated with
Arp~220. From the apparent magnitudes of the galaxies seen at this
position in optical survey plates, a redshift for the putative group can
be estimated, and the implied luminosity, size and temperature of the
southern feature turn out to be quite reasonable when compared to known
groups or poor clusters. Recent spectroscopic observations of two of
these galaxies do show them to have similar
redshifts ($z\approx0.09$; L.R.Jones; private
communication) comparable to the values estimated by Heckman \etal\ (1996).
This shows that there is at least a pair of galaxies behind the southern
feature, and strengthens the group interpretation.

In order to investigate the spectral characteristics of this southern
feature, a `southern' spectrum was extracted from a 1\arcm\ radius circle
centred at the position $\alpha=15^{h}34^{m}53.7^{s},
\delta=+23^{d}28^{m}32^{s}$. X-ray emission coincident with the
optical galaxies was excluded from the data to avoid possible
contamination of the spectrum from sources associated with the galaxies
themselves, and then standard spectral models were fitted. The model that
best fits this southern spectrum is a Raymond \& Smith (1977) hot plasma
model with an absorbing column (5.87$\pm2.74\times10^{20}$\,cm$^{-2}$)
consistent with that out of our Galaxy, a low temperature (0.5\,keV), and
a very low ($<0.05$ solar) metallicity. Further investigation of the
temperature-metallicity parameter space however, reveals that the
temperature is not very well constrained; the fit is only marginally worse
when the temperature is $\sim$1\,keV, and the metallicity is a few
tenths solar. At the distance of Arp~220, the best fit model,
once scaled to allow for the flux lost in the removal of emission
associated with the background galaxies, gives rise to an emitted
(0.1$-$2.0\,keV) X-ray luminosity of $1.4\times10^{41}$\,erg s$^{-1}$.
The equivalent luminosity at a distance corresponding to a redshift of
0.09 (360\,Mpc) is $2.9\times10^{42}$\,erg s$^{-1}$.

Assuming first that the southern feature {\em is} associated with
Arp~220, the total X-ray (0.1$-$2.0\,keV) luminosity is given by the
sum of the luminosities inferred from the spectral fits to each of the
`southern' and the `northern' spectra; $\log{L_{X}}=41.47$ (the same
figure is reached by the summation of the inferred luminosities of the
`diffuse' spectra and sources~A and B). If, however, the southern source is
{\em not} associated with Arp~220, then purely the `northern' spectrum
gives $\log{L_{X}}=41.18$. The true nature of this feature is discussed
later in Section~\ref{sec_disc_diff}.

\begin{figure} 
\vspace{80mm} 
\caption[]{\small
Contours of 0.1$-$2.3\,keV X-ray emission from Arp220 overlayed on an
optical image. The contour levels increase by factors of two from
$7.2\times10^{-4}$\,ct s$^{-1}$ arcmin$^{-2}$.} 
\label{fig_arp220}
\end{figure} 

\begin{table*}
\begin{center}
\begin{tabular}{cccccccccc}	\hline
\multicolumn{3}{c}{Source Spectrum Analysed} & Net  & $\log{L_{X}}$ &\multicolumn{5}{c}{Best
	Spectral Fit Results}  	\\ \cline{6-10}
    &R.A. & Dec &Counts& erg s$^{-1}$ & column & photon & temp. & Z
	& $\chi^{2}$ (n.d.o.f) \\ 
    &(2000.0)&(2000.0)&      &(0.1$-$2.0\,keV)& $10^{20}$\,cm$^{-2}$ &
	index & keV & (solar) & \\ \hline 
\multicolumn{3}{c}{`Northern'$^{a}$} & 168$\pm$18.7 & 41.18 &
	44.0$^{+19.5}_{-14.4}$ & & 0.21$^{+0.05}_{-0.04}$ & 0.00$^{+0.03}_{-0.00}$ 
	& 33.8 (18) \\ \\
\multicolumn{3}{c}{Diffuse$^{a}$}    & 220$\pm$33.4 & 41.37 &
   5.17$^{+2.46}_{-1.65}$ &  & 0.50$^{+0.13}_{-0.10}$ & 0.01$^{+0.02}_{-0.01}$ & \\ \\
Source~A & 15 34 54.31 & +23 28 27.0 & 49.1$\pm$9.5 & 40.38 &
	24.7$^{+46.7}_{-24.5}$ & 3.75$^{+2.85}_{-1.35}$ & & &  \\
Source~B & 15 34 56.91 & +23 30 10.6 & 107$\pm$12 & 40.55 &
	63.7$^{+35.3}_{-25.6}$ & & 0.23$^{+0.12}_{-0.07}$ & &  \\ \hline
\multicolumn{4}{c}{Diffuse + sources } & 41.47 &
& & & & \\ 
\hline
\end{tabular}
\caption[]{\small The results of the PSPC spectral analysis of Arp~220. 
Tabulated luminosities are those escaping from the system. Only the best model fit is shown.
Errors on the spectral fit parameters are 1$\sigma$ for one interesting parameter. \\
$^{a}$ The `northern' and diffuse spectra have been obtained in a different way from that
described in Section~\ref{sec_obs_data}. See text for details.} 
\label{results_arp220}
\end{center}
\end{table*}

	\subsection{NGC~2623}
	\label{sec_ngc2623}

NGC~2623 (Arp~243), eighth in the Toomre (1977) sequence of merging systems, exhibits two
large-scale, well defined tidal tails, blue colours and distinct regions of star formation
(Schombert \etal\ 1990). Both Toomre (1977) and Joseph and Wright (1985) infer from this
morphology that NGC~2623 is the product of a tidal encounter between two similarly massive disc
galaxies.

Whereas three distinct condensations can be identified in optical images of NGC~2623's central
region, near-infrared images (\eg Joy and Harvey 1987, Stanford and Bushouse 1991) reveal a
single symmetric nucleus lying between the two northern optical features. Joy and Harvey (1987)
conclude that the southern condensation cannot be a remnant nucleus, and may simply be a
counterarm generated by the tidal interaction (c.f. Toomre and Toomre 1972). Stanford and
Bushouse (1991) find that their K-band surface brightness profile is, as seen in Arp~220, 
well fitted by an
$r^{1/4}$ law, characteristic of elliptical galaxies, out to 3\,kpc. This was also seen by
Wright \etal\ (1990), who concluded that the stellar light profile of NGC~2623 bears a very
close resemblance to that of ellipticals.

A radio source in the nucleus of NGC~2623 was resolved at 6\,cm (Condon 1980), with a size of
170\,pc and a radio spectral index such that $F_{\nu} \propto \nu^{-0.8}$. The size of the
radio source, the slope of the radio continuum, and the ratio of 100\,$\mu$m to 21\,cm flux
(Joy and Harvey 1987) are consistent with synchrotron emission from supernovae produced in a
burst of star formation (Condon 1980, Condon and Broderick 1986). The radio observations also
appear to be consistent with a weak non-thermal emission-line source (c.f. Keel 1986).

Strong Balmer absorption lines reveal the presence of early-type stars in
NGC~2623's nuclear region, making up between one-third to one-half of the
central optical emission (Joy and Harvey 1987). Emission lines such as
H$\alpha$ and N {\scriptsize II}, in contrast to the early-type star
distribution, are found to originate only within the very centre of
NGC~2623, and the high N {\scriptsize II}/H$\alpha$ ratio observed
indicates that the nuclear emission lines are wholly or partially excited
by shocks or a weak nonthermal LINER source (Keel 1986). Joy and Harvey
(1987) conclude that, since nebular emission lines are not associated
with the extended distribution of young early-type stars, most of the
highly ionizing O and early B type stars have evolved off the main
sequence. A similar conclusion was reached by Larson and Tinsley (1978),
who found that NGC~2623 exhibits UBV colours of a galaxy in which a burst
of star formation occurred $\approx 10^{8}$ years ago.

There are several observations indicating that collisions between gas clouds are taking place
within NGC~2623. Joy and Harvey (1987), for instance, have inferred that the merger should
produce a region of compressed gas and dust $\sim2''$ in extent. This is observed within the
central regions of NGC~2623. Furthermore, observed strong OH absorption features imply the
existence of a dense concentration of molecular clouds in the nuclear region (Baan \etal\
1985). Also, CO observations (Sanders \etal\ 1986, Casoli \etal\ 1988) indicate that NGC~2623
has $6\times10^{9}$ M$_{\odot}$ of molecular hydrogen gas, highly concentrated towards the
radio and infrared nucleus, in a region less than 3\,kpc in size.

Once the analysis of the system had begun, exactly as described in
Section~\ref{sec_obs_data}, it soon became evident that only a small number of
counts were present. The analysis was therefore restricted to the energy band
(0.1$-$1.2\,keV), in order to maximise the signal to noise.
Figure~\ref{fig_ngc2623} shows contours of X-ray emission from NGC~2623 not only
in the `full' (here, 0.1$-$1.2\,keV) energy band, but also in soft
(0.1$-$0.5\,keV) and hard (0.5$-$1.2\,keV) bands. What is very evident from
figure~\ref{fig_ngc2623}, is that, as seen in Arp~220, two very distinct components
comprise the X-ray emission from NGC~2623; a hard, compact nuclear feature, and a
cool feature, lying outside of the optical confines of the system, above the
northern edge of the western tail.

\begin{figure} 
\vspace{120mm} 
\caption[]{\small
Contours of X-ray emission in three different energy bands (top: 0.1$-$1.2\,keV, 
middle: 0.1$-$0.5\,keV, bottom: 0.5$-$1.2\,keV) are shown superimposed on an
optical image of NGC~2623. The contour levels increase by factors of two from
$7.2\times10^{-4}$\,ct s$^{-1}$ arcmin$^{-2}$.} 
\label{fig_ngc2623}
\end{figure} 

Spectral fitting of the integrated, the diffuse and the nucear feature's spectra
all proved difficult, due to the small number of counts. Only very simple models
(thermal bremsstrahlung and power law with the absorbing column frozen at the
Galactic value) could be used to fit these spectra (see
Table~\ref{results_ngc2623}), and the maximum likelihood statistic had to be used
in every case. Thermal bremsstrahlung models proved the better fit to each
spectrum. Although Source~A, the nuclear feature, was not actually formally
detected at the 4$\sigma$ significance level, we have included it here on the
basis of it having a signal-to-noise ratio greater than 2. Furthermore, because of
the low number of counts, we have not given any spectral fitting results regarding
source~A. Source~A's spectrum does appear to be hard however (as indicated by
Figure~\ref{fig_ngc2623}), and a luminosity is given in
Table~\ref{results_ngc2623}, assuming a 1.5\,keV bremsstrahlung spectrum absorbed
by the column out of our Galaxy.

\begin{table*}
\begin{center}
\begin{tabular}{cccccccccc}	\hline
\multicolumn{3}{c}{Source Spectrum Analysed} & Net  & $\log{L_{X}}$ &\multicolumn{5}{c}{Best
	Spectral Fit Results}  	\\ \cline{6-10}
    &R.A. & Dec &Counts& erg s$^{-1}$ & column & photon & temp. & Z
	& $\chi^{2}$ (n.d.o.f) \\ 
    &(2000.0)&(2000.0)&      &(0.1$-$2.0\,keV)& $10^{20}$\,cm$^{-2}$ &
	index & keV & (solar) & \\ \hline 
\multicolumn{3}{c}{Integrated} & 36.0$\pm$10.3 & 40.93 & 3.28(F) & & 0.49$^{+0.28}_{-0.22}$ & & \\ \\
\multicolumn{3}{c}{Diffuse}    & 19.2$\pm$ 9.4 & 40.57 & 3.28(F) & & 0.20$^{+0.10}_{-0.08}$ &  & \\ \\
Source~A$^{a}$ 
   & 08 38 23.54 & +25 45 09   &  9.6$\pm$ 3.8 & 40.70 & 3.28(F) & & 1.50(F) & &  \\ \hline
\multicolumn{4}{c}{Diffuse + sources } & 40.94 & 
& & & & \\ 
\hline
\end{tabular}
\caption[]{\small The results of the PSPC spectral analysis of NGC~2623.
Tabulated luminosities are those escaping from the system. Only the best model fit is shown.
Errors on the spectral fit parameters are 1$\sigma$ for one interesting parameter. 
\\
$^{a}$ No spectral analysis results are given for source~A due to the low number of 
counts. The tabulated luminosity assumes a 1.5\,keV bremsstrahlung spectrum
corrected for Galactic absorption (see text).} 
\label{results_ngc2623}
\end{center}
\end{table*}

The X-ray emission from NGC~2623 is split into two very distinct components. The
nuclear feature (source~A) is compact, and strong, and appears to be spectrally
hard. It is coincident with published optical, radio, IR and CO emission. It is
also seen with the HRI (see Fig.~\ref{fig_ngc2623_hri}), though only at a very low
flux level ($3.6\times10^{-4}$\,cts s$^{-1}$).

The second feature, lying to the west of the system and coincident with seemingly
nothing whatsoever, is of much more interest. Comparison of the radial profile
with the PSPC point spread function is inconclusive as to whether the source is
extended or not. If the feature is point-like, then the very low temperature could
point to a serendipitous quasar or white dwarf as possible candidates. Typical
quasar spectra ($\alpha=2.2\pm$0.2; Branduardi-Raymont \etal\ 1994; Roche \etal\
1995) however, observed through the Galactic column to NGC~2623, can be shown
through simulations, to be far too hard to correspond to this soft component.
Furthermore, as only a few hundred white dwarfs were detected in the entire \Ros\
all-sky survey, it is statistically very unlikely that this feature is due to a
foreground white dwarf. It is also very unlikely that the soft feature is
associated with the western tail, as it lies well above it, and this tail is both
dimmer and less blue than its partner (Schombert \etal\ 1990). The HRI image
(Fig.~\ref{fig_ngc2623_hri}) shows something very interesting however as regards
this western feature -- nothing at all. Nothing is seen in the HRI data to the
west of NGC~2623 on any spatial smoothing scale down to a significance level of
2$\sigma$. The lack of any HRI emission is useful, as regards ascertaining whether
this feature is point-like or not. If the source were point-like then we should
have expected (based on the PSPC count rate) $\approx0.8$~ct ksec$^{-1}$ in the
HRI data. This is not seen; a $2\sigma$ upper limit to the HRI count rate at the
position of the PSPC feature gives $0.2$~ct ksec$^{-1}$. Hence, either the western
feature is very variable, or it is extended, on a scale substantially larger than
the spatial resolution of the HRI. We assume below that the western feature has a
{\em maximum} extent of $\approx2'$, as suggested by Figure~\ref{fig_ngc2623}.

\begin{figure} 
\vspace{80mm} 
\caption[]{\small
Contours of \Ros\ HRI X-ray emission from NGC~2623 overlayed on an
optical image. The contour levels increase by factors of two from
$2.2\times10^{-3}$\,ct s$^{-1}$ arcmin$^{-2}$.} 
\label{fig_ngc2623_hri}
\end{figure}

The fact that this feature is very soft is reminiscent of the X-ray
outflows (or winds) seen around many starburst galaxies (Heckman 1993;
Read \etal\ 1997). If NGC~2623's soft feature is an outflow, then it is a
rather different outflow from the standard starburst type; apart from it
being far more luminous and cooler, it appears only on one side of the
system, and it is a great deal larger ($\sim30$\,kpc).

This large cool, diffuse X-ray structure bears much more resemblance in fact, to the
feature seen in the previous section; the large amorphous structure to the south of
Arp~220. a system very similar to NGC~2623 (hence their adjacent positions in the
present sequence), with comparable IR luminosities, and similar elliptical-like light
profiles (Wright \etal\ 1990). The feature in Arp~220, as discussed in the previous
section, is believed by Heckman \etal\ (1996) to be associated with a galaxy group or
small cluster. In the case of NGC~2623, however, nothing is seen in the optical. The
nature of these features is discussed later in Section~\ref{sec_disc_diff}.

	\subsection{NGC~7252}
	\label{sec_ngc7252}

In Toomre's (1977) sequence of merging systems, NGC~7252 (Arp~226)
appears last, its structure, combining a single relaxed-looking body with
long, faint tidal tails, suggestive of a possible end-product of the
merger process -- an elliptical galaxy. Schweizer (1978) found many
features within this galaxy, often compared to the `atoms for peace'
symbol, which conclusively defined NGC~7252 as a late-stage merger. These
features included two similar, oppositely-moving tidal tails, a
single relaxed nucleus, chaotic motions within the main body, ripples and
loops at the edges of the main body, and no nearby
perturbing systems. Further work showed that the light from NGC~7252
follows an $r^{1/4}$ law (\eg Stanford \& Bushouse 1991), typical of
ellipticals, it exhibits a post-starburst nuclear population (Schweizer
1982), and it obeys the Faber-Jackson relationship (Lake \& Dressler
1986), again characteristic of normal elliptical galaxies. Many more
recent studies have strengthened the idea that NGC~7252 is a late-stage
merger of two gas-rich spiral discs (\eg Schweizer 1990; Hibbard \etal\
1994), and it is now considered to be the prototype merging galaxy
remnant. Perhaps the strongest observational support of the merger
hypothesis comes from the recent VLA observations of NGC~7252 (Hibbard
\etal\ 1994), where, while the tidal tails were seen to be very abundant
($2\times10^{9} h^{-2} M_{\odot}$) in atomic hydrogen, a fact requiring a
gas-rich spiral genesis, the main body of the system was seen to be
virtually empty of \hi, and appears indistinguishable from a typical
elliptical. 
The molecular gas, in contrast to the atomic gas, is tightly
bound to an inner rotating disc, of radius $\approx$7\arcs, the gap between
the atomic and molecular gas distributions apparently filled by diffuse 
H${\alpha}$ emission. 
The \hi\ within the tidal tails of NGC~7252 is
very interesting: the 
$M_{\hi}/L_{B}$ ratio is consistent with that seen in the outer regions of
late-type spirals ($\sim1$; Wevers \etal\ 1986), and the gas and 
starlight in the eastern tail appear to have the same extent, whereas in
the northwestern tail, the neutral hydrogen appears to extend some
40\,kpc or so beyond the optical emission. From the
work of Cayatte \etal\ (1994), this may indicate that
the progenitor galaxy that gave rise
to the eastern tail may be of an earlier type than its partner (Hibbard
\etal\ 1994).

Computer simulations of collisions between two gas-rich discs (Borne \&
Richstone 1991; Mihos
\etal\ 1993) reproduced the essential features of NGC~7252's appearance at first, quite
adequately, though a retrograde encounter geometry (\ie galaxy rotations
in the opposite sense to their mutual orbit), and a very tightly bound
orbit, such that the galaxies were in contact at the start of the
simulation, were necessary. These simulations were later however, found
to be inconsistent with the \hi\ data of Hibbard \etal\ (1994), and only
the more recent simulations of Hibbard \& Mihos (1995) are able to match
the observations (and with a much more astrophysically satisfying orbital
geometry), explaining the \hi\ in terms of the material within NGC~7252's
tails, having reached a turning point in its expansion, falling back to
smaller radii. It is thought that approximately half of the present tail
material ($\approx10^{9} M_{\odot}$ of \hi, and $2\times10^{9} L_{\odot}$
of starlight) will return to the main body over a timescale of less than
1 Gyr.

NGC~7252 has never been observed in X-rays previously, and the \Ros\ PSPC and HRI observations
presented here, give us a first glimpse at this important system in X-rays.
A brief analysis of the PSPC data has already been published in Hibbard \etal\ (1994), and their
X-ray image appears very similar to our Figure~\ref{fig_ngc7252}, though we have contoured the
present image to a greater depth. 
The X-ray source coincident with NGC~7252 appears very centrally concentrated and compact,
though a couple of tentative extended features are seen, apparently connected to the main body,
to the east and to the northeast. A comparison of the PSPC PSF with a radial profile of the
X-ray data centred on the peak of the emission, suggests that the source is resolved, though
only at the 2$\sigma$ significance level.

The best fit to the integrated spectrum indicates that the emission from NGC~7252 may be due
mainly to quite cool (0.43~\,keV) gas. As discussed in the last paragraph, the emission from
NGC~7252 may only just be resolved, and consequently, the source-subtraction procedure
described in Section~\ref{sec_obs_data} had great difficulties in splitting up this emission
into source and diffuse emission. In fact, only when a very large radius (4\arcm) was used, 
could the fitting of the diffuse spectrum settle at sensible values. 
As can be seen from Figure~\ref{fig_ngc7252} however, the use of such a large extraction
radius merely results in the inclusion of emission from features having nothing to do with
NGC~7252, notably from features along the left-hand edge of Figure~\ref{fig_ngc7252} (see
Section~\ref{sec_tails}).

Consequently, we have assumed the integrated spectral fit results to be the best indicator of
the true (0.1$-$2.0\,keV) X-ray luminosity; $5.3\times10^{40}$\,erg s$^{-1}$. The spectrum
is best fitted by a low temperature plasma, indicative of hot gas enveloping the system
(the fact that the fitted absorbing column appears larger than the Galactic value probably
indicates some contribution from harder, unresolved sources, embedded deeper within the galaxy).

\begin{figure} 
\vspace{80mm} 
\caption[]{\small
Contours of 0.1$-$2.3\,keV X-ray emission from NGC~7252 overlayed on an
optical image. The contour levels increase by factors of two from
$7.2\times10^{-4}$\,ct s$^{-1}$ arcmin$^{-2}$.} 
\label{fig_ngc7252}
\end{figure}

\begin{table*}
\begin{center}
\begin{tabular}{cccccccccc}	\hline
\multicolumn{3}{c}{Source Spectrum Analysed} & Net  & $\log{L_{X}}$ &\multicolumn{5}{c}{Best
	Spectral Fit Results}  	\\ \cline{6-10}
    &R.A. & Dec &Counts& erg s$^{-1}$ & column & photon & temp. & Z
	& $\chi^{2}$ (n.d.o.f) \\ 
    &(2000.0)&(2000.0)&      &(0.1$-$2.0\,keV)& $10^{20}$\,cm$^{-2}$ &
	index & keV & (solar) & \\ \hline 
\multicolumn{3}{c}{Integrated} & 225$\pm$17.2 & 40.73 & 5.20$^{+2.68}_{-2.68}$ &
& 0.43$^{+0.21}_{-0.21}$ & & 17.6 (19) \\ \\
\multicolumn{3}{c}{Diffuse$^{a}$}    & 142$\pm$46.2 & 40.39 & 0.00$^{+1.83}_{-0.00}$ & 
 & 0.09$^{+0.05}_{-0.09}$ & 0.00$^{+0.32}_{-0.00}$ & \\ \\
Source~A & 22 20 44.54 & -24 40 43.9 & 165$\pm$14.3 & 40.65 &
	6.09$^{+1.79}_{-1.39}$ & & 0.49$^{+0.14}_{-0.10}$ & &  \\ \hline
\multicolumn{4}{c}{Diffuse + sources } & 40.84 &
& & & & \\ 
\hline
\end{tabular}
\caption[]{\small The results of the PSPC spectral analysis of NGC~7252.
Tabulated luminosities are those escaping from the system. Only the best model fit is shown.
Errors on the spectral fit parameters are 1$\sigma$ for one interesting parameter.\\
$^{a}$ The diffuse spectrum has been obtained in a different way from that
described in Section~\ref{sec_obs_data}. See text for details.} 
\label{results_ngc7252}
\end{center}
\end{table*}

A number of these sources in fact appear in the HRI image
(Figure~\ref{fig_ngc7252_hri}), where a great deal of structure is seen. Two
sources are formally detected close to the central body of the system (see
Table~\ref{results_ngc7252_hri}). Source~H1 is bright, and appears to be
coincident with the nucleus, lying less than 7\arcs\ from the position given in
the RC3. Features are also seen almost symmetrically opposed to the southeast and
northwest of the nucleus. Source~H2 is especially interesting, as firstly, it
appears to be coincident with the base of the eastern extension tentatively seen
in the PSPC image (Figure~\ref{fig_ngc7252}), and secondly, it is the only
significant feature detected in the hard band. Interestingly, the northeastern
feature seen in the PSPC image appears also to have a tentative counterpart
visible in the HRI image (to the top-left of Figure~\ref{fig_ngc7252_hri}).

\begin{figure} 
\vspace{80mm} 
\caption[]{\small
Contours of \Ros\ HRI X-ray emission from NGC~7252 overlayed on an
optical image. The contour levels increase by factors of two from
$2.2\times10^{-3}$\,ct s$^{-1}$ arcmin$^{-2}$.} 
\label{fig_ngc7252_hri}
\end{figure}

\begin{table}
\begin{center}
\begin{tabular}{ccccc}	\hline
       & R.A. & Dec. & Net & $\log{L_{X}}$ \\
       & (2000.0) & (2000.0) & counts & (erg s$^{-1}$) \\ \hline
H1     & 22 20 44.56 & -24 40 36.4 & 19.9$\pm$5.3 & 40.83 (40.67) \\
H2     & 22 20 46.56 & -24 40 49.2 & 6.82$\pm$3.1 & 40.36 (40.21) \\
\hline
\end{tabular}
\caption[]{\small Sources within NGC~7252 detected with the HRI. Luminosities are calculated 
as described in Section~\ref{sec_hri}, using a 0.3\,keV (and a 3.0\,keV) thermal 
bremsstrahlung model, corrected for Galactic absorption. } 
\label{results_ngc7252_hri}
\end{center}
\end{table}

	\subsection{AM~1146-270}
	\label{sec_am1146-270}

AM1146-270 was selected from the Arp-Madore {\em Catalogue of Southern Peculiar Galaxies and
Associations} (Arp and Madore 1987). It has a very complex knotty structure within its central
$10''$, as indicated by the B-band image of Smith and Hintzen (1991), where seven local maxima,
none of which seem to be superimposed stars, are seen. A plume or fan ($13''$ in length)
extends from the western side of the galaxy (see Fig.~\ref{mergers_pics}). The colours are very
blue, similar to those of an irregular galaxy, and many blue knots are seen surrounding the
nucleus (Smith and Hintzen 1991). This galaxy was detected by IRAS and appears to be the site
of much recent star formation.

AM1146-270 received only 601 seconds of PSPC time. No significant
emission was detected from this system, so a 2$\sigma$ upper limit to the
count rate was derived at the position given in Smith and Hintzen (1991);
4.47\,ct ksec$^{-1}$. Assuming a 1\,keV bremsstrahlung plasma and Galactic
absorption ($5.18\times10^{20}$\,cm$^{-2}$) then an upper limit to its emitted
(0.1$-$2.0\,keV) X-ray luminosity can be derived;
$L_{X}<3.9\times10^{39}$\,erg s$^{-1}$ (the equivalent 5\,keV bremsstrahlung
upper limit is $4.1\times10^{39}$\,erg s$^{-1}$).

\section{Discussion $-$ The X-ray Evolution of Merging Galaxies}
\label{sec_discussion}

The previous subsections have described the X-ray properties of eight systems thought to be at
different evolutionary stages of a merger of two disc galaxies. These systems range from
`approaching' mergers, such as Arp~270, and Arp~242, where the discs of the two galaxies are
still quite distinct, and bridges and tails have only just begun to form, through to `contact'
mergers, such as the {\em Antennae} and NGC~520, where, though the galactic nuclei are still
quite distinct, the galactic discs have begun to interact strongly, through to `ultraluminous'
mergers, such as Arp~220 and NGC~2623, extremely powerful systems on the point of nuclear
coalescence, and finally, to `remnant' mergers, such as NGC~7252 and AM~1146-270, relaxed,
elliptical-like systems, their violent star-formation episodes long-since over.

Although we are assuming here that these eight systems represent an
evolutionary sequence, one must be careful not to assume that they
represent eight different evolutionary stages of the same system, for
several reasons. Firstly, the progenitor galaxies are/were all different
in detail, in terms of masses, mass ratios, gas richness \etc\ Secondly,
the process by which each system has come about, \ie the encounter
geometries and parameters, will have all been different. Both of these
facts are bound to affect the evolution of the system. The fact however,
that we have used the appearance, length and symmetry of each system's
tidal tails as one of our main selection criteria, means that we are
likely to have selected broadly similar systems, \ie prograde,
medium impact parameter, collisions of similarly-sized disc galaxies.
Forces need to act for a long time for the formation of strong tails, and
this requires a predominantly prograde encounter (Toomre \& Toomre 1972).
Similarly, the force needs to be strong, \ie the impact parameter needs
to be small (Barnes 1992), though not so small as to cause the galaxies
to merge before the tails have had a chance to grow. Lastly, proper,
symmetrical tail construction requires that the two masses should be
roughly equal (Toomre \& Toomre 1972).

Therefore, we should really think of the following evolutionary discussion in conservative
terms, not as snapshots of a single system at different points in its evolution, more in
general terms of how systems appear compared to one another, at `approaching', `contact',
`ultraluminous', and `remnant' stages. 

\subsection{Previous related work} 
\label{sec_previous_work}

Before continuing, it is useful to review here more fully relevant
work on the evolution of interacting systems in other wavebands.

Joseph \& Wright (1985) studied the infrared properties of a sample of
merely intermediate-age mergers (\ie covering a smaller evolutionary period
than in the present study) including three of the systems within our
sample, namely NGC~520, NGC~2623, and Arp~220. None of the systems
covered either showed two distinct discs, as in the `nearby' mergers, or
appeared as relaxed or elliptical-like as NGC~7252. Nevertheless,
their results are of great importance; they observed the infrared
luminosity to rise and fall along their sequence, suggesting
that a massive burst of star-formation of luminosity $\sim10^{12}
L_{\odot}$ occurs within merging galaxies as the merger progresses.

Telesco \etal\ (1988) studied a large sample
of comparable-sized interacting galaxies from the Arp-Madore {\em Catalogue of Southern
Peculiar Galaxies and Associations} (Arp and Madore 1987). They found that the dust colour
temperature and the interaction strength are both at their largest for systems with the
smallest galactic separations. This result, along with those of Joseph \& Wright (1985) above,
have, as discussed in Section~\ref{sec_sample}, been used in our selection and chronological
ordering criteria (see Table~\ref{table_sample}).

More recently, perhaps the most important work concerning the evolution
of merging galaxies has been the work of Hibbard \& van Gorkom (1996),
and it is worth summarising some of their more important results here, as
many of them will have a substantial bearing on the discussion of the
X-ray results in the next section.

Hibbard \& van Gorkom (1996) presented aperture synthesis observations of
the neutral hydrogen distribution in a sample of five systems (four of
which are taken from the Toomre (1977) sequence, and three of which
(Arp~242, NGC~520 and NGC~7252) appear here). This they supplemented with
wide field, narrow band H$\alpha$ images and deep R-band surface
photometry.

They found that, within the inner regions, the starlight in the later
systems (including NGC~7252) exhibited $r^{1/4}$ profiles, a natural
consequence of violent relaxation (Lynden-Bell 1967; van Albada 1982),
and obeyed the Faber-Jackson relationship, indicating that as these later
systems can be dated at about 1~Gyr after the merger (Schweizer 1996),
the primary characteristics of ellipticals are created rather rapidly
compared to the total merging time-scales. 

The ionized hydrogen is seen, in the early systems (including Arp~242),
to form minor axis plumes, similar to the wind `blow-outs' of Heckman
\etal\ (1993), and in the later systems, to be confined to the inner kpc,
presumably associated with a small molecular gas disc.

The cold gas is especially interesting, and is seen to be widely
separated in the later systems, with the molecular gas concentrated within the
remnant bodies and the atomic gas predominantly in the outer regions.
Simulations have shown that the symmetric forces experienced
during a merger can force at most half of the outer disc material
into a tail, the rest of the \hi\ being forced into the inner regions
(Hibbard \& Mihos 1995). As less than a quarter of the total \hi\ is
found within these regions however, Hibbard \& van Gorkom (1966)
concluded that most of the centrally forced gas must have been converted
during the merger to some other form.

One possibility for what has happened to this gas, is that it has been
turned into stars, and indeed, the presence of an `E$+$A' spectrum
(Dressler \& Gunn 1983) in NGC~7252 (Fritze-von Alvensleben \& Gerhard
1994), and in the almost identical NGC~3921 (Schweizer 1996), indicates
that a significant amount of the original atomic gas has formed
stars. Another possibility for the fate of the atomic gas, is that it has
been converted into molecular gas, though the fact that both NGC~7252 and
NGC~3921 have below average molecular gas contents for their spiral
progenitors (Solomon \& Sage 1988; Young \& Knezec 1989), indicates that
there has been little net conversion of atomic to molecular gas. A third
possibility for the lost atomic gas, is that it has been heated to X-ray
temperatures, either through cloud-cloud collisions (Harwitt \etal\
1987), or through the energy input from massive stars and supernovae
created in one (or, as discussed later, more than one) starburst(s). This
neatly brings us on to the X-ray evolution.

\subsection{The X-ray evolution}
\label{sec_evolution}

In Table~\ref{table_results} we have gathered together the emitted
0.1$-$2.0\,keV X-ray luminosities of the eight sample systems,
calculated as explained in Sections~\ref{sec_arp270} to
\ref{sec_am1146-270}. Note that in the case of Arp~220, we have given the X-ray
luminosity of only the northern feature, the southern outflow/group
emission has been omitted (inclusion of this emission raises
$L_{X}$ to 2.9$\times10^{41}$\,erg s$^{-1}$ -- see
Section~\ref{sec_arp220}).

Also shown in Table\,\ref{table_results}, are the statistical errors
(expressed as percentages), based on the number of counts (including
background), for each source.
There is, in addition to this statistical error, a second error which
arises from uncertainty in the appropriate model to use. Since the
X-ray luminosities derived here are band limited to the 0.1$-$2.0\,keV
band, the effects of considering different acceptable models are not very
large. Comparing the luminosities inferred from power-law and hot plasma
models in cases where these give fits of comparable quality, we find
differences of $\sim$1-15\% (the larger variations applying to the lowest
count-rate cases). For a given spectral model, the effects of varying the
model parameters within the 90\% confidence contour results in luminosity
variations of the same order. As can be seen from
Table~\ref{table_results}, these uncertainties in $L_{X}$ are similar in
size to the simple statistical flux errors.

\begin{table}
\begin{center}
\begin{tabular}{|l|l|l}	\hline
System &Other names & $\log{L_{X}}$   \\ 
       &            & (erg s$^{-1}$) \\ 
       &            & (\% error)     \\ \hline
Arp~270    &NGC~3395/6& 40.49$^{a}$   (10.3 \%) \\ 
Arp~242    &NGC~4676  & 40.76$^{b}$   (35.3 \%) \\ 
NGC~4038/9 &Arp~244   & 41.03$^{a,c}$ ( 2.6 \%) \\ 
NGC~520    &Arp~157   & 39.94$^{b}$   ( 9.5 \%) \\ 
Arp~220    &UGC~9913  & 41.18$^{b,d}$ (11.1 \%) \\ 
NGC~2623   &Arp~243   & 40.93$^{b}$   (28.6 \%) \\ 
NGC~7252   &Arp~226   & 40.67$^{b}$   ( 7.6 \%) \\ 
AM~1146-270&          & $<$39.61$^{e}$  (-)     \\ 
\hline
\end{tabular}
\caption[]{\small The merging galaxy sample. Emitted (0.1$-$2.0\,keV)
X-ray luminosities have been calculated as described in
Sections~\ref{sec_arp270} to \ref{sec_am1146-270}, by using the spectral
fits to various spectra, as follows (quoted errors are purely
statistical; see text):\\
$^{a}$ Summation of the best 2-component fit to the diffuse spectrum plus fits to 
individual source spectra \\
$^{b}$ Best fit to the integrated spectrum. \\
$^{c}$ See Read \etal\ (1995). \\
$^{d}$ Note for Arp~220, the `northern' spectrum has been used, 
\ie emission from the southern
component has been ignored (inclusion of this component increases $L_{X}$ to
2.9$\times10^{41}$\,erg s$^{-1}$).\\
$^{e}$ 2$\sigma$ upper limit (using a 5\,keV bremsstrahlung model).} 
\label{table_results}
\end{center}
\end{table}

In order to see how these X-ray luminosities 
compare with the various multiwavelength
properties given in Table~\ref{table_sample}, what is needed is a plot of
the various different luminosity ratios ($L_{X}/L_{B}$, $L_{X}/L_{FIR}$
\etc) as a function of merger age. As discussed earlier, even
establishing a reliable chronological ordering is fraught with
difficulties, and dynamical ages for these
systems are considerably more uncertain. However, we are interested in
broad trends through the merger process, rather than precise timescales.

With this in mind, Fig.~\ref{fig_evolution} shows the temporal variation
of the X-ray, the far-infrared, the optical and the radio luminosities
for the merging galaxy sample. Also plotted is the variation of the
far-infrared colour temperature, $S_{60}/S_{100}$. All values have been
scaled to those of the first system in the sequence, Arp~270. Shown
for comparison, again scaled to the values of Arp~270, are, to the left,
the mean values of $L_{FIR}/L_{B}$, $L_{X}/L_{B}$ and $L_{X}/L_{FIR}$ of
both the starburst and normal spiral galaxy samples from Read \etal\
(1997), and to the right, the sample of elliptical and S0 galaxies from Mackie \&
Fabbiano (1997). 
When compared to the results of Fabbiano \etal\ (1992), we find
that the X-ray luminosities of both NGC~7252 and AM~1146-270 (upper limit) are very
low when compared to typical ellipticals. It should be noted however, that
$L_{X}/L_{B}$ for ellipticals is very $L_{B}$-dependent, and so, the `remnant
stage' galaxies' X-ray properties should really be compared with those of
elliptical galaxies of similar optical luminosity. It is seen though, that the
X-ray luminosity of NGC~7252 does appear to be quite low when compared to similarly
optically bright systems (again from Fabbiano \etal\ 1992), and although only an
upper limit to the X-ray luminosity from AM~1146-270 could be established, this
limit is still quite restrictive, and its X-ray luminosity also appears to be
quite low when compared to systems with a similar (very low) optical luminosity 
(note that very few systems appear in this region of 
$L_{X}-L_{B}$ parameter space in Fabbiano \etal (1992),
so this result is not very secure).

The interaction timescales in Fig.~\ref{fig_evolution} are certainly
only approximate, though in the
establishing the evolutionary positioning, we have been able to use
a variety of pointers. Firstly, for convenience, we have placed time zero
at the point of nuclear coalescence, chosen here to be at the position of
the brightest system Arp~220. Now, it is thought from its tail kinematics
and spectral properties (Fritze-v.\,Alvensleben \& Gerhard 1994;
Schweizer 1990) that NGC~7252 is at $\sim$1\,Gyr after nuclear merger
-- hence
its positioning. AM~1146-270's colours and morphology suggest that it is
somewhat older. On the younger side of the evolution, comparison of the
general morphologies of the sample galaxies with the numerical
simulations of Mihos \& Hernquist (1996) suggest $\sim$650 Myr between
an Arp~270-type system and nuclear coalescence. Finally, NGC~520, whose
simulation lookalikes seem to occur over halfway in, is thought to have
experienced extranuclear starburst activity, as is ongoing in the {\em
Antennae}, and perhaps the {\em Mice}, some 300\,Myr ago (Stanford 1991).

\begin{figure*}
\vspace{140mm}
\caption[]{\small The evolution of the merging galaxy sample. Plotted is the temporal
variation of the X-ray, the far-infrared, the optical and the radio
luminosities for the merging galaxy sample. Also shown is the variation
of the far-infrared colour temperature, $S_{60}/S_{100}$. All values have
been scaled to those of the first system in the sequence, Arp~270. Also
shown for comparison, are, to the left, the mean values of
$L_{FIR}/L_{B}$, $L_{X}/L_{B}$ and $L_{X}/L_{FIR}$ of both the starburst
(stars) and normal (diamonds) spiral galaxy samples of Read \etal\ (1997),
and to the right, the Mackie \& Fabbiano (1997) elliptical and S0 sample.}
\label{fig_evolution}
\end{figure*}

The evolution appears to start, as one would perhaps expect, at an
activity level somewhere between that of normal and starburst spiral
galaxies. The level of activity, whether given by $L_{FIR}/L_{B}$,
$L_{X}/L_{B}$, $L_{RAD}/L_{B}$ or $S_{60}/S_{100}$, rises during the
evolution, to well in excess of that of typical starburst galaxies, and
then falls again, back to a very similar level to that at the start.
Bearing in mind that elliptical galaxies may well be the end-products of
this evolution, it is interesting to note that the final $L_{X}/L_{B}$
level is quite low when compared to that of typical ellipticals, though
it is within the accepted activity range. Interestingly, there is
some  evidence that the continued evolution of $L_{X}/L_{B}$ of
even later merger remnants (\ie\ moving further to the right) is seen to
rise gradually to a level in line with that of typical ellipticals
(Mackie \& Fabbiano 1997). The X-ray properties of the these merger
remnants are discussed in more detail in Section~\ref{sec_final_stages}.

What may be of critical importance is a comparison of the far-infrared
evolution (which we may assume bears a very close resemblance to the
evolution of the current star-formation rate, SFR) with the theoretical
work of Mihos \& Hernquist (1996). These authors used numerical
simulationsto investigate the gas dynamics 
and starbursts in mergers similar to those
considered here (\ie of systems of comparable mass). They found that the
actual structure of the galaxies plays a dominant role in the evolution
of the SFR. Two merging scenarios, identical except for
the make-up of the systems, were considered. One consisted of a merger of pure
disc/halo systems, systems completely lacking a bulge component, the
other consisted of a merger of disc/bulge/halo systems, each dense bulge
component having a mass some three times that of the disc component. It
was found that the gaseous inflows (which are related to the SFR, and
hence to the FIR luminosity) are strongest when galaxies with dense
central bulges are in the final stages of merging. 

Firstly, let us discuss the SFR evolution of the non-bulge mergers. The
SFR in these systems is seen to peak at 40$-$60 times field spiral values
at a time some 500 Myr before nuclear coalescence, this peak lasting for
about 150 Myr, \ie from a time around the position of Arp~242, till soon
after the position of the {\em Antennae}. As these starbursts deplete the
nuclear regions of gas, the starbursts die out, and the SFR is seen to
drop again close to its quiescent value at about the position of NGC~520.
During nuclear coalescence (at time zero), only a relatively weak
starburst is produced (an increase in the SFR by about an order of
magnitude, and lasting for around 50-100 Myr). The SFR history of the
bulge mergers however, is very different and bears a very close
resemblance to Fig.~\ref{fig_evolution}. The starbursts that develop
after the initial encounter are much weaker than those seen in the
bulgeless case. The SFR is increased by only a factor of a few (less than
an order of magnitude), though, as the gas is not significantly depleted,
it lasts longer, from about Arp~270, or just after, to just before
Arp~220. Consequently, at close to time zero, far more gas exists, and
the rapid collapse of this gas leads to a very violent starburst, with an
SFR some 150 times that of a field spiral, lasting for only 50\,Myr. In
both the bulge and the non-bulge cases, continued infall of the tidal
debris is unable to fuel any significant star-formation, and the remnant
after nuclear merger evolves passively at a SFR level 
similar to that of field spirals.

In reality, galaxies may have a wide range of bulge-to-disc ratios,
and mergers occur between galaxies of all
types. Although our merging sample has not been chosen with a constant
bulge-disc ratio in mind, all the systems seem to show the signature of
high-bulge progenitors, when compared to Mihos \& Hernquist's (1996)
work. The early systems (Arp~270, Arp~242 and the {\em Antennae}) show
only modest increases in $L_{FIR}$, perhaps by a factor of ten, above
normal field spirals. Similarly, the `ultraluminous' systems' progenitor
galaxies must have had large bulge-to-disc ratios, because of their
massive inferred SFRs. This does throw up further caveats regarding the
difficulty of an evolutionary analysis. Firstly a disturbed morphology
does not necessarily mean that a very large burst of star-formation is
taking place, and secondly, the main requirements for the existence of an
ultraluminous starburst may be specific encounter geometries and/or
specific galaxy structures (Hibbard \& van Gorkom 1996).

Comparing the X-ray and far-infrared evolution, one sees that although
both rise and fall along the merger sequence, the X-ray does so to
nowhere near the extent of the far-infrared (and, to a lesser extent, the
radio). This leads to the drop by well over an order of magnitude in
$L_{X}/L_{FIR}$ in Fig~\ref{fig_evolution}. Before concentrating on this
important feature of the evolution however, the striking `glitch' in
the $L_{X}/L_{B}$ evolution associated with the system NGC~520, should
first be discussed.

NGC~520, the `second brightest very disturbed galaxy in the sky' (Arp
1987), while once thought, as discussed in Section~\ref{sec_ngc520}, to
be a single exploding galaxy (\eg Arp 1967), is now believed, after
receiving so much attention over the years, to be a merger-induced system
(\eg Stanford \& Balcells 1990, 1991; Bernl\"ohr 1993). Its far-infrared
properties, as seen in Fig.~\ref{fig_evolution}, appear to sit nicely on
the evolutionary sequence. Its X-ray properties however, do not. NGC~520
is very underluminous in the X-ray, with an $L_{X}/L_{B}$ ratio similar
to that of normal (\ie\ non-starburst) field spirals. Note that NGC~520's
very low $L_{X}/L_{FIR}$ ratio (similar to that of Arp~220 and NGC~2623)
is very unlikely to be due to the same phenomenon as in the cases of
these ultraluminous mergers, as they have {\em high} $L_{X}/L_{B}$
ratios, NGC~520 does not. It is, quite simply, X-ray dim.

This can be seen very clearly in Fig.~\ref{fig_lx_lfir}, where the X-ray
and far-infrared luminosities of the eight merging galaxies are plotted
(AM~1146-270's upper limit as an arrow), together, for comparison, with
Read \etal's (1997) samples of normal and starburst spiral galaxies. Best
fit regression lines (excluding the AM~1146-270 upper limit) to the
merging galaxy sample data are plotted both inclusive and exclusive of
NGC~520. Also plotted is the best fit regression line to the Read \etal\
(1997) sample.

\begin{figure*}
\vspace{80mm}
\caption[]{\small Plot showing the \Ros\ (0.1$-$2.0\,keV) X-ray luminosity versus far-infrared
luminosity, $L_{FIR}$, calculated as described in Section~\ref{sec_sample}, for all members of
the merging galaxy sample (filled squares) (AM~1146-270's upper limit is shown by the arrow).
Also shown, for comparison, are the results for the samples of normal (diamonds) and starburst 
(stars) spiral galaxies from Read \etal\ (1997). Best fit regression lines (excluding the
AM~1146-270 upper limit) to the merging galaxy sample data are plotted both inclusive (dashed
line) and exclusive (solid line) of NGC~520. Also plotted is the best fit regression line to 
the Read \etal\ (1997) sample.}
\label{fig_lx_lfir}
\end{figure*}

NGC~520 stands out quite obviously in Fig.~\ref{fig_lx_lfir}. Although
very far-infrared bright, as bright as the brightest non-interacting
starbursts, it is X-ray underluminous by nearly an order of
magnitude. Why? There is quite a lot of evidence to suggest that NGC~520
is a quite different system from the rest of this merging sample.

A strange feature of NGC~520 is the fact that the disc is still
remarkably intact; a rotating disc of neutral hydrogen of mass a few
$\times10^{9} M_{\odot}$ and radius $\sim20$\,kpc is seen within the
centre of NGC~520 (Hibbard \& van Gorkom 1996). Optical isophotes
indicate that both nuclei are embedded within this disc. The disc has
survived the merger so far very well. What is to say it that it will not
survive the nuclear coalescence? Why NGC~520's disc has survived so well
is somewhat of a mystery, though numerical simulations (\eg Quinn \etal
1993) suggest that the mass ratio may be very important. For
dynamically cold discs to be significantly disturbed, violently
fluctuating fields such as those accompanying moderate mass-ratio
interactions ($M_{1}/M_{2}>0.3$) are required. 
For NGC~520, whilst molecular linewidth
studies (Sanders \etal\ 1988a) and stellar velocity dispersion
measurements (Stanford \& Balcells 1990) suggest mass ratios close to
unity, rotation curves (Bernl\"ohr 1993) give a mass ratio of 10:1.

While the presence of a large gaseous disc within the
merging system requires at least one of the progenitors to have been gas
rich, Hibbard \& van Gorkom (1996) observe the extensive plumes within
NGC~520 to be gas-poor ($M_{\mbox{\hi}}/L_{R}\lapx 0.1$), suggesting that
NGC~520 came about through the merger of gas-poor progenitor discs
(S0$-$Sa galaxies). 
The most likely scenario offered by Hibbard \& van Gorkom is
that the NGC~520 system is a result of an encounter between a gas-rich
system with an extensive disc and a gas-poor system, such as an S0 or Sa
galaxy. They also conclude though that NGC~520 shows strong evidence that
mergers may not necessarily destroy the gaseous discs of the progenitors
and therefore need not evolve into ellipticals. Another related
possibility, given that extensive-disc (\ie near-bulgeless) systems may
be involved, is that NGC~520 may be the result of a merger between two
bulgeless systems, \ie as in the evolution of Mihos \& Hernquist (1996)
discussed earlier, though its relatively low SFR would mean that it must be
at an evolutionary stage sometime after the early strong starburst seen
in these simulations. Either way, because of the different encounter
progenitors involved, whether in terms of mass ratio, gas richness or
bulge-to-disc ratio, NGC~520 does appear to be on a significantly
different evolutionary path to the rest of the sample.

Turning now to the X-ray evolution of the sample as a whole, it is worth
studying Fig.~\ref{fig_evolution} and Fig.~\ref{fig_lx_lfir}
together. A rather tight $L_{X}$:$L_{FIR}$ relationship is seen
in Fig.~\ref{fig_lx_lfir}, running quite smoothly and continuously from
quiescent normal spirals (diamonds), through starbursts (stars) to very
active merging galaxies (filled squares). There does appear however, to
be a gradual flattening of the slope at higher activities, and this is
borne out by the large change in the slopes of the regression fits, from
0.80, for normal spirals, to 0.38 (0.31 when NGC~520 is discounted) for
the merger sample.

Why does this flattening of the $L_{X}$:$L_{FIR}$ relationship occur?
Also, why does the large drop in the X-ray to far-infrared lumnosity
ratio, seen in Fig~\ref{fig_evolution}, occur?. We believe the answer to
both questions to be the same, as both the $L_{X}$:$L_{FIR}$ flattening,
and the $L_{X}/L_{FIR}$ drop are really just the same effect manifesting
itself in two different ways. As the merger progresses, starburst
activity is seen to rise and fall. The increases in both the far-infrared
and the X-ray luminosity are both due to the increased starburst
activity. While the infrared flux is primarily due to dust heated by the
massive stars within the starburst, the rise in 
X-ray flux is limited, as much of
the input energy from the large number of supernovae and stellar winds
created within the starburst, is lost. It is most likely that this energy
is lost in the form of kinetic energy -- kinetic energy associated with
the huge gaseous ejections seen in the X-ray in the most luminous members
of the merging galaxy sample. We now move on to discuss these diffuse
structures.

\subsection{Diffuse structures}
\label{sec_disc_diff}

Diffuse structures, as described in the notes on the individual systems,
are observed throughout almost the whole of the evolutionary sequence.
Arp~270's two point sources appear to be embedded in a small amount of
diffuse emission, though this emission does not seem to extend very far
(if at all) beyond the optical confines of the system. The two sources
are both best-fitted by warm plasma models, however, suggestive of
hot gas. Arp~242's distance makes it hard to make any firm judgements
about the extent of any diffuse structures. The fact that the source
appears extended however, along with the fact that the emission is
significantly softer than in Arp~270, is very suggestive of significant
amounts of hot gas. Furthermore vigorous disc-wide star-formation is seen
in H$\alpha$ within this system, along with plumes of gas extending along
the minor axis, possibly coincident with the north-west X-ray feature.
The coincidence of H$\alpha$ and X-ray features extending along the minor
axis of this system points towards the existence of a galactic wind or
`blowout' (\eg Heckman 1993), NGC~4038/9, as discussed briefly here, and
in detail in Read \etal\ (1995), shows a great deal ($\sim10^{9}
M_{\odot}$) of low temperature (0.4\,keV) gas, not only enveloping the
two discs, in what appears to be a hot corona, but also apparently
ejected from the system to the north and south as a pair of galactic
winds. NGC~520 is very compact, and shows little evidence for any
significant amount of diffuse gas. However, as we have already discussed,
this system may well be an oddity. Arp~220, the most
active of the sample, shows a 40\,kpc long, east-west structure,
coincident with a `double-bubble' H$\alpha$ emission-line nebula
(Heckman, Armus \& Miley 1987;1990). This X-ray/optical emission is
generally believed to be due to a bipolar wind driven out from the
nucleus by an ultrapowerful starburst. The large soft source to the south
of the galaxy is apparently connected to Arp~220, though it may in fact
be associated instead with a background galaxy group, as discussed in
Section~\ref{sec_arp220}. However,
NGC~2623, a very similar system to
Arp~220, exhibits a near-identical feature, which seems a remarkable
coincidence. NGC~2623's very soft ($\sim0.2$\,keV) feature
appears to be truly diffuse, given that it is not seen with the HRI.
NGC~7252, a near elliptical-like remnant, shows some evidence for an
X-ray halo. This halo of hot gas surrounding NGC~7252, however, appears
small. And finally, only an upper limit could be calculated for
AM~1146-270, due to the very short observing time. It is however, quite a
restrictive limit, and places the activity level of AM~1146-270 close to
that of NGC~7252, and definitely far less than that of the ultraluminous
systems.

The time evolution of the temperature, the size, and the
$L_{X}(\mbox{diff})/L_{B}$ values of these diffuse structures are
summarized in Fig.~\ref{fig_evol_diff}. All values have been scaled to
those of the {\em Antennae} since Arp~270's diffuse emission is only
suggestive. Two sets of points are plotted for Arp~220, one (the higher
set) for the case where the southern source {\em is} assumed to be
associated with Arp~220, and the other (the lower set), for the case
where it is not. In a discussion of the trends seen within this figure,
one must be very conservative. The extraction of truly diffuse emission
and the fitting of models to these spectra has been, as described
earlier, difficult. Even so, there is evidence that, as a merger
progresses, hot gas is ejected from the system. This gas is seen to grow
both in size and in X-ray
brightness. Curiously, by the late stages of the evolution, there
is little evidence for much gas. Little can be said about the temperature
evolution, except that it appears to remain low ($<0.5$\,keV) throughout
the merger.

\begin{figure*} 
\vspace{140mm} 
\caption[]{\small The evolution of the diffuse structures seen 
within the merging galaxy sample. Plotted are the temporal variations of the temperature
(stars), the size (in terms of the maximum extent) (squares) and the values of
$L_{X}(\mbox{diff})/L_{B}$ (diamonds). All values have been scaled to those of the {\em
Antennae}. Upper limits are given by arrows. Two sets of points are given for Arp~220, one
(the higher set) for the case where the southern source {\em is} assumed to be associated with
Arp~220, one (the lower set), for the case where it is not. In the case of NGC~520, it proved
impossible to extract any believable information regarding any diffuse component, and no
values are plotted.} 
\label{fig_evol_diff} 
\end{figure*}

A major problem with forming ellipticals from spirals is to account for
the disappearance of the gas from the original spirals. Star formation is
too inefficient a process to suppose that it is all converted into stars,
and so the bulk is probably blown away by supernova explosions and
stellar winds, and ultimately, galactic winds. We believe that we are
seeing this ejected gas in X-rays, notably in the {\em Antennae}, Arp~220
and NGC~2623.

A great range of gaseous structures are observed within the sequence, and
three fundamental (and chronological) questions need to be addressed;
When and where does this gas ejection process start? What is the true
nature of the huge one-sided structures seen in the ultraluminous sytems?
Why is there only marginal evidence for a hot halo in the remnants?

\subsubsection{Early stages}
\label{sec_disc_diff_early}

Even in a pre-merger system such as the {\em Antennae}, a great deal of
hot gas is seen, some of it being forcibly ejected from the system. What
about still earlier systems? One would perhaps intuitively think that no
ejection could occur before the galaxies first encounter each other (\ie before
Arp~270), and this idea is strengthened by the modelling work of Mihos \&
Hernquist (1996) in which no significant star-formation
occurs before merging galaxies first meet. Arp~270 appears to be at such
an early stage in its encounter, it just may be on the verge of gaseous
ejection. Remember that this system is at such an early starburst episode
that, although star-formation is taking place at a level between that of
typical normal and starburst spirals, it may be occurring well away from
the nuclei, the gas not having had time yet to be compressed
into the nuclear regions. Arp~242, at a later stage, is almost certainly
ejecting hot gas into its halo. Its distance makes it difficult for us to
say for sure, but its X-ray structure, and temperature, along with the
coincidences of $H{\alpha}$ features, point towards `blowout'-type
ejection taking place within this system.

\subsubsection{Mid-stages}
\label{sec_disc_diff_mid}

These ejections of hot gas are likely to grow as the merger progresses,
as an increasing amount of
gas is compressed into the nuclei, giving rise to more
and more supernova explosions. How then do these gaseous structures
appear at the peak of merger-induced activity? Arp~220 and NGC~2623 are
very similar systems. Both are extremely bright, especially in the
infrared and in the radio, and are members of the class of ultraluminous
far-infrared galaxies (FIRGs) discovered by IRAS (Sanders \etal\ 1988b).
Both appear to be at an evolutionary stage corresponding to the short
(50$-$100\,Myr) violent starburst period, seen in the Mihos \& Hernquist
(1996) simulations. At first glance, both appear very similar in the
X-ray, with a large, soft, gaseous feature
apparently connected to the galaxy, but situated some way from it, and a
harder, more compact source centred on the galaxy itself. Furthermore,
the same phenomenon is seen in \Ros\ observations of a third violently
merging, ultraluminous FIR galaxy, NGC~6240 (Fricke \& Papaderos 1996).
Again, in this galaxy, which is very similar to both Arp~220 and NGC~2623
in terms of its extreme luminosity and position in the merging
evolution, a massive ($\sim50$\,kpc), one-sided, diffuse feature is seen
in the soft band image, whereas in the hard band image, only a point-like
nuclear source is visible.

The similarities between these three systems strongly suggest that the
large soft features are associated with the merging galaxies. On the
other hand, as discussed in Section~\ref{sec_arp220}, there is evidence
for a group of galaxies at $z\approx 0.09$ lying behind the soft feature
in Arp~220. The properties of the diffuse X-ray emission from
galaxy groups have been established over the past few years
as a result of studies with \Ros\ (Mulchaey \etal\ 1996, Ponman
\etal\ 1996). The temperature of the hot gas is typically $\sim 1$\,keV,
and the luminosity (for $H_{0}=$75\,km s$^{-1}$
Mpc$^{-1}$) ranges widely from
$10^{40}$\,erg\,s$^{-1}$ to $10^{43}$\,erg\,s$^{-1}$, with a strong positive
correlation with temperature. Unfortunately, as reported in
Section~\ref{sec_arp220}, the spectral properties of the diffuse source in
Arp~220 are not very well constrained. At the best fit temperature of
$\sim$0.5\,keV, the derived luminosity of $2.9\times10^{42}$\,erg\,s$^{-1}$
would be very high for a group, but if $T\sim1$\,keV (as is allowed by the
data) it would be quite reasonable. Hence the group hypothesis for the
soft source cannot be ruled out by the available data.

We have a choice, then, between two coincidences. Either the source to the
south of Arp~220 is a background group, in which case it is a remarkable
coincidence that very similar features (with no sign of any background
galaxies) are seen in the two similar systems NGC~2623 and NGC~6240,
or the source is associated with Arp~220 itself, in which case
(unless we are prepared to invoke non-cosmological redshifts) it must
be a coincidence that there is a group, or at least a pair, of galaxies
behind it.

One might hope, if these features are indeed galactic outflows, that
one would see some corresponding emission from them at other wavelengths.
In the case of NGC~2623, the soft X-ray source extends well
outside the optical galaxy, but the H$\alpha$+N{\scriptsize II} emission-line
image (Armus, Heckman and Miley 1990) shows very little structure in the
direction of this outflow. It is believed that both soft X-ray emission 
and optical line emission are produced by the shocking of cool clouds
embedded in a hot wind, and most normal systems where hot outflows are
believed to exist (\eg NGC~253, M82) show X-ray and
optical line emission features of reasonably similar size
(see Heckman \etal\ 1993). Arp~220 is a
case in point, its optical emission line halo, as discussed in
Section~\ref{sec_arp220}, is approximately coincident with the smaller,
east-west X-ray halo. Nothing is seen however, coincident with the huge
southern extension.

What are these features? If
they truly are hot gas ejected from the merger, then several questions
arise. Why do they have no counterparts at other wavelengths?
Why are they one-sided?

The Mihos \& Hernquist (1996) simulations suggest that in a
merger between two similar-sized, gas-rich, large-bulge systems (as we
believe we have sampled here), {\em two}, quite different bursts of
star-formation occur. The first occurs as the sytems first meet each
other (\ie from Arp~270 to somewhere close to the Arp~220 epoch), and is
of a comparatively low level (a level, in fact, similar to that of nearby
starbursts, such as NGC~253 and M82), and lasts for a comparatively long
time. The second starburst is far more intense, and lasts for an
extremely short time ($\sim50$\,Myr), occurring around the time of
nuclear coalescence (our time zero), \ie at the Arp~220/NGC~2623 epoch.
This starburst is very different from those experienced by the earlier
`approaching' and `contact' mergers.

Consider now the possible evolution. The first starburst, because of its
similarity to M82-types, is likely to produce similar bipolar outflows,
as are observed, both in the {\em Antennae} (Read \etal\ 1995), and
in Arp~220 (Section~\ref{sec_arp220}; Heckman \etal\ 1996).
Interestingly, however, in both cases, there is evidence that the `bipolar'
nature of these outflows has become distorted by the evolution of the
merger (as discussed in Sections~\ref{sec_ngc4038} and \ref{sec_arp220},
and in Read \etal\ (1995) and Heckman \etal\ 1996)). An alternative
explanation for their appearance may relate to the creation of
`unipolar' outflows, as discussed below.

Perhaps the strangest facet of the large, soft features seen in the most
lumunious systems is their
unimodality. Why do the structures seen in Arp~220, NGC~2623 (and
NGC~6240) appear only on one side of the system? The answer may lie in
the fact that these systems are evolving rapidly. Mac Low \& McCray
(1988), using the Kompaneets (1960) (thin-shell) approximation,
numerically modelled the growth of superbubbles: large thin shells of
cold gas surrounding a hot pressurized interior -- the progenitors
of galactic winds -- in various stratified atmospheres.
Among their findings, they discovered that superbubbles blow out of the
\hi\ layer  when they grow to a radius of between one and
two scaleheights. The swept up shell then accelerates outwards and
fragments, due to Rayleigh-Taylor instabilities, and the wind is then
able to escape into the inter-galactic medium at velocities of several
thousand km s$^{-1}$ (see Heckman \etal\ 1993).

However, Mac Low \& McCray (1988) also find that these bubbles will blow
out on {\em one side only} of a disc galaxy if their centres are more
than 50$-$60\,pc from the centre of the disc. Now, in a (relatively)
dynamically stable system, such as M82 or NGC~253, the starburst is found to be
very symmetrically positioned with respect to the galactic disc, and
bipolar structures are seen in the X-ray (Watson, Stanger \& Griffiths
1984; Fabbiano 1988; Pietsch 1992; Strickland, Ponman \& Stevens 1997;
Read \etal\ 1997). In a rapidly-evolving ultraluminous merging system
such as Arp220 or NGC~2623 however, the central burst is highly unlikely
to be so centrally positioned with respect to the quickly-moving and
highly distorted gaseous components. There will then
be a {\em single} direction of steepest
pressure gradient, along which the bubble will expand most rapidly,
leading to a one-sided blowout. 

It is also possible that this might apply to earlier systems
within the present sample, as alluded to above. The distorted
`bipolar' structure seen in the {\em Antennae}, for instance
(Section~\ref{sec_ngc4038}; Read \etal\ 1995), may not be a bipolar wind
after all, but rather a pair of unipolar winds, one produced from
the northern starburst, one from the southern. There is some evidence to
support this, both in terms of the fact that the individual discs have
become very distorted, and that the individual nuclear starbursts appear
to be offset (see Read \etal\ 1995). The same could be also true in the
case of the `double bubble' structure seen in Arp~220. Two individual
unipolar winds, as opposed to one bipolar wind may have produced this
structure.

Mean physical properties for these hot gas structures can be inferred
from our spectral fits if we make some assumptions about the geometries
of the emission. We have assumed that both the Arp~220 southern feature
and the NGC~2623 feature are spherical in nature, on the basis of the
fact that both the Arp~220 and the NGC~6240 feature (Fricke \& Papaderos
1996) appear quite round (the NGC~2623 feature appears quite round also,
though we are unsure whether it has been resolved with the PSPC
(see Section~\ref{sec_ngc2623})). This is the simplest assumption to make
and indicates the largest volume over which the gas is likely to be
distributed. Using the volumes derived for this `bubble' model, the
fitted emission measure $\eta n_{e}^{2} V$ (where $\eta$ is the `filling
factor' -- the fraction of the total volume $V$ which is occupied by the
emitting gas) can be used to infer the mean electron density, $n_{e}$,
and hence the total mass $M_{\mbox{\small gas}}$, thermal energy
$E_{\mbox{\small th}}$ and cooling time $t_{\mbox{\small cool}}$ of the
gas, and also the mass cooling rate $\dot{M}_{\mbox{\small cool}}$ and
adiabatic expansion timescale $t_{\mbox{\small exp}}$. These values for
the Arp~220 southern feature and the NGC~2623 eastern feature are given
in Table\,\ref{table_diff_param}, along with, for comparison, the
corresponding values for the {\em Antennae} corona (Read \etal\ 1995),
and for the NGC~7252, assuming that entire integrated emission from
NGC~7252 is in the form of a hot gaseous halo.

\begin{table*}
\begin{center}
\begin{tabular}{|l|c|c|c|c|}    \hline
 & NGC~4038/9 & Arp~220 & NGC~2623 & NGC~7252 \\ \hline
$\log{L_{X}}$ (erg s$^{-1}$) & 40.61 & 41.15 & 40.57 & 40.73  \\
$T_{X}$ (keV) & 0.36 & 0.50 & 0.20 & 0.43 \\
$n_{e}$ (cm$^{-3}$) & $0.0049\times1/\sqrt{\eta}$&
    $0.0047\times1/\sqrt{\eta}$ &
    $0.0069\times1/\sqrt{\eta}$ & $0.0105\times1/\sqrt{\eta}$ \\
$M_{\mbox{\small gas}}$ ($M_{\odot}$) & $1.5\times10^{9}\sqrt{\eta}$
    & $8.5\times10^{9}\sqrt{\eta}$
    & $2.4\times10^{9}\sqrt{\eta}$ & $1.8\times10^{9}\sqrt{\eta}$ \\
$E_{\mbox{\small th}}$ (erg) & $3.1\times10^{57}\sqrt{\eta}$
    & $2.5\times10^{58}\sqrt{\eta}$
    & $2.7\times10^{57}\sqrt{\eta}$
    & $4.4\times10^{57}\sqrt{\eta}$ \\
$t_{\mbox{\small cool}}$ (Myr) & $2410\sqrt{\eta}$
    & $5590\sqrt{\eta}$ & $2290\sqrt{\eta}$
    & $2590\sqrt{\eta}$ \\
$\dot{M}_{\mbox{\small cool}}$ ($M_{\odot}$ yr$^{-1}$)    & 0.62 & 1.52
    &1.06& 0.70 \\
$t_{\mbox{\small exp}}$ (Myr) & 59 & 90 &82 & 38 \\
\hline
\end{tabular}
\caption[]{\small Values of physical parameters for the diffuse gas (using a hemispherical
bubble model; see text) seen in the NGC~4038/9 corona, the Arp~220 southern feature, the 
NGC~2623 eastern feature, and the NGC~7252 halo. $\eta$ is the
filling factor of the gas.}
\label{table_diff_param}
\end{center}
\end{table*}

If, as we are assuming, we are seeing some sort of outflow within these ultraluminous
systems, then if $\eta\approx1$, they are presumably bubbles of gas at
2$-6\times10^{6}$\,K flowing out from the system. If this is the case however, this gas,
unless some additional force is confining it, should expand at its sound speed (a few
hundred km s$^{-1}$), which is too slow when compared to the size of these features
($\approx50$\,kpc)and the timescale of the starburst ($\approx50$\,Myr). Alternatively,
the soft X-ray emission we see could arise from material, \eg clouds, shock heated by a
fast wind (see Heckman \etal\ 1993). 
Although this hypothesis is supported by recent theoretical modelling of starburst winds
(\eg Suckov \etal\ 1994), the actual value of the filling factor of the gas
emitting in the \ros\ band, remains unclear, some simulations (\eg Suchkov \etal\ 1994) 
predicting a low value, others (\eg D.K.Strickland, private communication) predicting a 
high value. In the low filling factor case, the \ros\ band X-ray luminosity of the clouds
should far exceed that of the wind fluid itself. In the high filling factor case, the 
wind fluid contibutes a very significant fraction to the \ros\ band emission. 

In contrast to the corona surrounding the {\em Antennae} galaxies, therefore, which
appears to be bound to the system ($\eta\approx1$; see Read \etal\ 1995), the offset
features seen in Arp~220 and NGC~2623 are (like the {\em Antennae} outflows) likely to
be due to a combination of dense clouds, shocked by a hot, fast unbound wind, and the
wind fluid itself. Provided that this material does escape from the system, then the
associated mass loss rates can be calculated from the inferred gas masses, if we assume
a velocity for the material. This gives 140 (for Arp~220) and 50 (for NGC~2623)
$\times\sqrt{\eta}v_{1000}$\,$M_{\odot}$ yr$^{-1}$, where $v_{1000}$ is the velocity, in
units of 1000\,km s$^{-1}$. Similarly, the rate at which kinetic energy is lost can be
estimated to be 4.8 (Arp~220) and 1.7 (NGC~2623)
$\times10^{43}\sqrt{\eta}v_{1000}^{3}$\,erg s$^{-1}$.

{\em If} $\eta$ were $\approx1$, \ie if we were seeing the hot wind
itself, rather than shocked clouds with $\eta\ll1$, then this mass-loss
could remove a substantial amount of gas ($\sim10^{10}$\,$M_{\odot}$)
from the merger (a requirement, if these objects are to eventually
resemble gas-poor ellipticals), even over the short starburst timescale
of $50-100$\,My. Furthermore, the rate at which kinetic energy could be
lost turns out to be quite comparable to the total inferred supernova
luminosities of these systems (calculated using the method of Read \&
Ponman (1995)); 28 and 6 $\times10^{43}$\,erg s$^{-1}$, respectively for
Arp~220 and NGC~2623. Hence, {\em if} $\eta$ were $\approx1$, then almost
all of the mass and around a quarter of the energy injected into these
systems by supernovae could escape in the wind.

If $\eta$ is substantially less than unity, the above results still stand, though they
refer to the mass and energy lost in the clouds (provided, of course, they escape the
system). This reduces the mass and energy loss rates calculated, but there is of course
additional mass and energy loss associated with the wind fluid itself, which is
invisible to \Ros. Simulations (Suchkov \etal\ 1994; D.K.Strickland, private
communication) appear to show though, that the lower density phases of the gas do not 
carry a significant fraction of the mass. 

It is also interesting to compare the values calculated above with the 
total energy and mass deposition rates, calculated (as in Heckman \etal\ 1993)
in terms of the starburst bolometric luminosity $L_{\rm bol}$ (which in the 
cases of Arp~220 and NGC~2623 we may assume is approximately equal to $L_{\rm
FIR}$). The energy deposition rates $\dot{E} = 8\times10^{42} L_{\rm
bol,11}$\,erg s$^{-1}$ (where $L_{\rm bol,11}$ is in units of $10^{11}
L_{\odot}$), are 6.9 (Arp~220) and 1.5 (NGC~2623)
$\times10^{43}$\,erg s$^{-1}$, and the mass deposition rates $\dot{M} = 3 L_{\rm
bol,11}$\,$M_{\odot}$ yr$^{-1}$, are 25.9 (Arp~220) and 5.8
(NGC~2623) $M_{\odot}$ yr$^{-1}$. 

As a final point, note that in the above discussion, we have only
considered the Arp~220 southern feature, not the northern `double-bubble'
emission. Inclusion of this emission into the analysis, \ie performing
the same calculations to the results of fitting the {\em total} `diffuse'
spectrum (see Section~\ref{sec_arp220}), gives rise to estimated gas
masses and thermal energies almost a factor of two greater than those
given in Table~\ref{table_diff_param}. Hence if this material is also
escaping, it will approximately double the mass and energy loss
rates discussed above.

\subsection{Final stages} 
\label{sec_final_stages}

The X-ray properties of the end products of our evolution, NGC~7252 and
AM~1146-270, are rather intriguing (even though we were only able to obtain an
upper limit for AM~1146-270, it is still quite a restrictive limit). As discussed
earlier, the X-ray luminosities of these systems are low compared to ellipticals
in general, more consistent with field spirals than with ellipticals. Furthermore, 
when compared to ellipticals of the same optical luminosity, it appears that both
galaxies may be  relatively X-ray dim.
This, combined with the fact that the \Ros\ PSPC image of NGC~7252
(Fig.~\ref{fig_ngc7252}) shows little in the way of any extended
halo emission, poses a problem.

In the previous two subsections, we have discussed the massive amounts of
hot gas seen in the vicinity of these earlier interacting systems. The gas which
is present in winds will leave the system on a short timescale, so the
fact that these X-ray structures are not seen at the NGC~7252 epoch, a
Gyr or so later, is not too surprising.
The gaseous corona seen in the {\em Antennae} however, is different
(coronal structures such as that within the {\em Antennae} may exist
within Arp~220 and NGC~2623 as well, though limitations on spatial
resolution and on numbers of counts make it impossible to say). The {\em
Antennae}'s coronal gas, given the fact that it appears largely bound,
with $\eta$ close to unity (Read \etal\ 1995), is likely to have a long
cooling time, perhaps 1$-$2\,Gyr, and hence such a gaseous halo should
remain until the NGC~7252 epoch. This starburst generated gas may be the
origin, as discussed earlier, of the hot gaseous halos known to exist
around many large ellipticals (\eg Fabbiano 1989). For this hypothesis to
work, one would expect hot gaseous halos around merger remnants. Why
then, given that there is a good deal of evidence that hot gas should
remain around the system until after obvious morphological evidence of a
merger has vanished, does NGC~7252, the prototypical merger remnant
candidate, show little sign of a hot halo?

The problem is not in reconciling the observations of the
earlier systems with those of NGC~7252. As can be seen from
Table~\ref{table_diff_param}, the coronae seen in the {\em Antennae} and 
in NGC~7252 are very similar in terms of amount of gas. It is true
though that, as the {\em Antennae} merger progresses through the
ultraluminous phase, more energy and mass is likely to be injected into
the halo, making it uncomfortably large when compared to the
NGC~7252 X-ray observations. On the other hand
cooling is taking place, returning the hot coronal gas to the disc, and
over the 1.5\,Gyr or so between the {\em Antennae} phase and the NGC~7252
phase, even at its present (relatively low) cooling rate, $10^{9}
M_{\odot}$ of material will have returned to the central system. The real
problem is that, in order for the merger hypothesis to work, and in order
for these merger remnants to resemble elliptical galaxies in the X-ray,
the halos need to be far larger ($\sim50-100$\,kpc), more massive
($\sim10^{9}-10^{10} M_{\odot}$; Fabbiano 1989; Forman
\etal\ 1993) and hotter ($T\approx0.7-1.2$\,keV; Matsushita \etal\ 1994; Rangarajan \etal\
1995) than that observed around the prototype remnant, NGC~7252.

The solution may be that we are looking for elliptical-like X-ray
properties in merger remnants of far too young a dynamical age. For
instance, it is thought that it will still take a few billion years for
the optical luminosity, brightness distribution, colour gradients,
velocity dispersion, UBV colours and fine structure (shells, loops,
ripples \etc) of NGC~7252 to evolve to that consistent with a normal
elliptical (Schweizer 1996). Perhaps it will take as long for the X-ray
properties to do the same.

In remants such as NGC~7252, quite a
substantial amount of gas still remains, both in the tidal debris and in
the central molecular disc. Hibbard \& van Gorkom (1996) believe that
NGC~7252 needs to get rid of half of its molecular gas and about 90\% of
its tidal material, in order to appear as an elliptical galaxy. The
removal of the central molecular gas appears to present no real problem.
What remains can be converted into stars within the next few Gyr, given
current central star-formation rates (1$-2 M_{\odot}$\,yr$^{-1}$; Hibbard
\etal\ 1994). The removal of the tidal debris however may be of
particular interest. In a recent N-body simulation of NGC~7252 (Hibbard
\& Mihos 1995), it was seen that the majority of the material within the
interaction-induced tidal tails is actually bound to the remnant. The
bases of the tails fall back quickly, whereas the more
distant material falls back more slowly. Some fraction of the material
however may remain far from the remnant in very long tails, for very long
times (\eg 20\% of the tidal material is believed to remain beyond
90\,kpc for over a Hubble time).  Although this material would be far too
faint to observe, it may contribute to the Ly$\alpha$ forest and to the
IGM (Hibbard \& van Gorkom 1996). This return of the tidal debris
therefore, can stretch out the merger remnant lifetime to many Gyr.

The fact that the atomic gas disappears rather abruptly at the
base of NGC~7252's northwestern tail, a region where the \hi\ gas is seen
to be falling toward smaller radii, whereas the optical tail continues
across the face of the galaxy, is very suggestive of a quick and efficient
process taking place converting the infalling gas to other phases
(Hibbard \etal\ 1994). One possibility as to what happens to the \hi\ gas
is that it is converted into molecular gas through compression, in which
case the system will remain rich in molecular gas for a long time. The
other possibility is far more attractive; this returning material may be
shock heated by the hot, though at this epoch, small halo. As this
process continues, the X-ray luminosity and size of this halo will grow.
The accompanying injection of energy will also raise the temperature. This is
perhaps how large, hot X-ray halos are formed around merger remnants, (a
requirement for the success of the merger hypothesis). The main bulk of
the halo is not formed through the ejection of hot gas from the merger
during the starburst stages of the interaction, but is instead formed,
much later, through the shock-heating of tidal tail material falling back
onto the remnant. If this picture is correct, then it is not
surprising that only a small X-ray halo exists around NGC~7252. There is
still a long time to wait (a few Gyr) before the bulk of the tail
material returns, forming the large X-ray halos seen around many
ellipticals.

\subsection{X-rays from tidal tails}
\label{sec_tails}

There is much evidence for enhanced star-formation, and even dwarf galaxy
formation, within merger induced tidal tails (\eg\ Schweizer 1978;
Schombert \etal\ 1990). It is known that large amounts of neutral
hydrogen are found within such tails (\eg\ van der Hulst 1979; Hibbard
\etal\ 1994), and the self-gravity of this material can cause clumping,
and lead eventually to star-formation. At the tip of the {\em Antennae}'s
southern tail, for instance, Mirabel, Dottori \& Lutz (1992) have
discovered a dwarf galaxy in the process of forming. This feature
however, a chain of nebulae ionized by very young ($<6$\,Myr) stars, is
found to emit no significant X-ray emission (Read \etal\ 1995). This is not
too surprising, given that, if all the stars have been formed within the
last 6\,Myr, even the most massive of these would not have had time to
evolve off the main sequence into the X-ray bright supernova remnant or
high-mass X-ray binary phase. In later systems, however, X-ray features
may be visible, as stars may have had time to evolve to X-ray bright
phases.

In practice, no significant emission is seen coincident with any tidal
tail features anywhere within our merging galaxy sample. The soft diffuse
feature seen to the west of NGC~2623, at first glance, may appear to have
something to do with the western tail. However, as discussed in
Section~\ref{sec_ngc2623}, this is very unlikely to be the case. Even in
NGC~7252, a system showing several large clumps of gas within its tidal
tails (Hibbard \etal\ 1994), no obvious X-ray counterparts are
observed. A small amount of emission is seen however, within the eastern
tail, spatially coincident with a giant \hii\ region (Hibbard \etal\
1994) -- the bluest region within the whole of NGC~7252.
This feature (the `plume' of emission to the north of the source at
$\alpha=22h20m56.25s$, $\delta=-24d42m06.4s$) contains only $\sim7$
counts however, and a 2$\sigma$ upper limit to its
(0.1$-$2.0\,keV) luminosity (assuming a 5\,keV bremsstrahlung model) is
$L_{X} < 1.7 \times10^{39}$\,erg s$^{-1}$. Converting the optical
luminosity of this \hii\ region (1.8$\times10^{41}$\,erg s$^{-1}$;
Hibbard \etal\ 1994) into an X-ray luminosity, assuming a similar
$L_{X}/L_{B}$ ratio to NGC~7252 as a whole, gives
$L_{X}=1.7\times10^{38}$\,erg s$^{-1}$, a factor of ten below our upper
limit. It is not, therefore, surprising that we do not detect any
significant X-ray emission from this region.

\section{Conclusions}
\label{sec_conclusions}

We have completed a programme of \Ros\ PSPC and HRI observations of a
carefully chosen chronological sequence of eight merging systems in order
to study the evolution of their X-ray properties through the merging
process. While we have discussed the X-ray emission from each individual
system in some detail, our main conclusions pertain to the evolutionary
sample as a whole, and these conclusions are outlined below:

\begin{itemize}

\item Across the sequence, from separated discs to relaxed remnants, the
activity level is seen to rise, from a level between that of nearby
normal and starburst spirals, to a level far in excess of that of nearby
starbursts, returning, at the end of the sequence to a level similar to
that at the start.

\item Large amounts of hot gas are seen being forcibly ejected from
these systems, the quantities of gas, and their extents, increasing with
activity. The gas ejection process starts soon after the
galaxies first meet each other, and evolves into what
appear to be bipolar `blowout'-type winds. These `bipolar' features may
be similar to those seen in nearby starbursts, the distortions due to the
rapid evolution of the systems, or they may be due to two separate
`unipolar' winds, one ejected from each of the two interacting galaxies.
What appear to be massive, unipolar, ejections of hot gas are observed at
the ultraluminous phases of the interaction, at the point of nuclear
coalescence. These may result from a far more violent starburst taking
place at this phase than at earlier stages, combined with very fast and
vigourous dynamical evolution of the system. At later times however, in the
elliptical-like, relaxed remnant phase, almost all of this hot gas
appears to have vanished, leaving only the suggestion of a small 
X-ray emitting halo.

\item The X-ray luminosity evolution, like the
general activity level, rises and falls along the evolutionary sequence.
However, the amplitude of this change is considerably (more than
an order of magnitude) smaller than that
seen in the far-infrared. The resulting large drop in the $L_{X}/L_{FIR}$
ratio is probably realted to the ejected hot gas. The FIR flux
is primarily due to dust heated by massive stars within the starburst,
and hence tracks the star formation history. However, the fraction of
this energy which is radiated by hot gas is reduced as the starburst
proceeds, as much of the input energy from supernovae
and stellar winds actually goes into the kinetic energy of the ejected
gas. 

\item The prototypical merger remnant NGC~7252, bears little resemblance
to typical elliptical galaxies in the X-ray, a fact at first appearing to
conflict with the `merger hypothesis', whereby elliptical galaxies can be
formed from the merger of two disc galaxies. One possible resolution of
this problem is that we are looking at far too early an evolutionary age.
Although NGC~7252's X-ray halo is far smaller, less massive, less bright
and cooler than those of typical ellipticals, shock heating of the
returning tidal debris over the next few Gyr, may increasing the size, mass,
luminosity and temperature of the halo. This may be the main source of an
eventual elliptical-like X-ray halo, rather than the remnants of hot gas
ejected into the halo at earlier phases.

\item The observations of NGC~520 reveal that it has no
discernable diffuse component, is very underluminous in the X-ray, and
sits very uncomfortably in the X-ray evolutionary sequence. It appears
from other multiwavelength studies, that the NGC~520 merger
has left the disc intact, whether because of gas-poor progenitors or a
low progenitor mass ratio, and is likely not to evolve into an
elliptical galaxy.

\end{itemize}

\section*{Acknowledgements}

AMR acknowledges the receipt of a SERC/PPARC studentship during the
initial phases of this work, and the Royal Society for more recent support. 
Ray Wolstencroft was involved in the early stages of this work, and we
would like to thank him and also Ian Stevens, Dave Strickland and Duncan 
Forbes 
for valuable discussions. We thank Laurence
Jones for optical observations of the {\em Antennae} and Arp~220
which have helped to clarify the nature of sources close to these
systems. AMR also also acknowledges colleagues from the MPE \Ros\ group
for their support. 
This research has made use of data obtained
from the UK \Ros\ Data Archive Centre at the Department of Physics and
Astronomy, Leicester University, UK.
We also thank the referee for useful comments which have improved the
paper.
Data reduction and analysis was performed on the Starlink node at
Birmingham.

\label{lastpage}

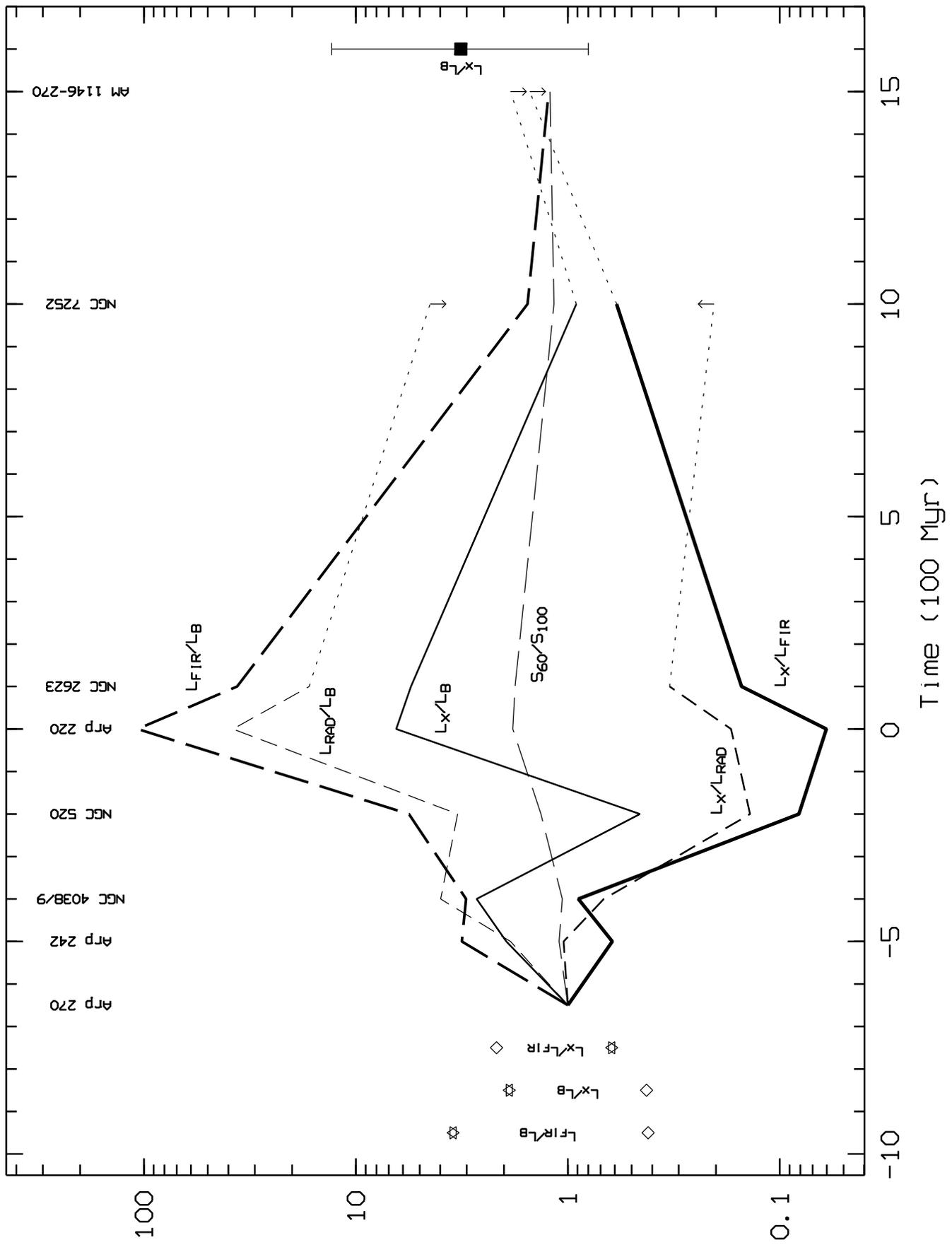
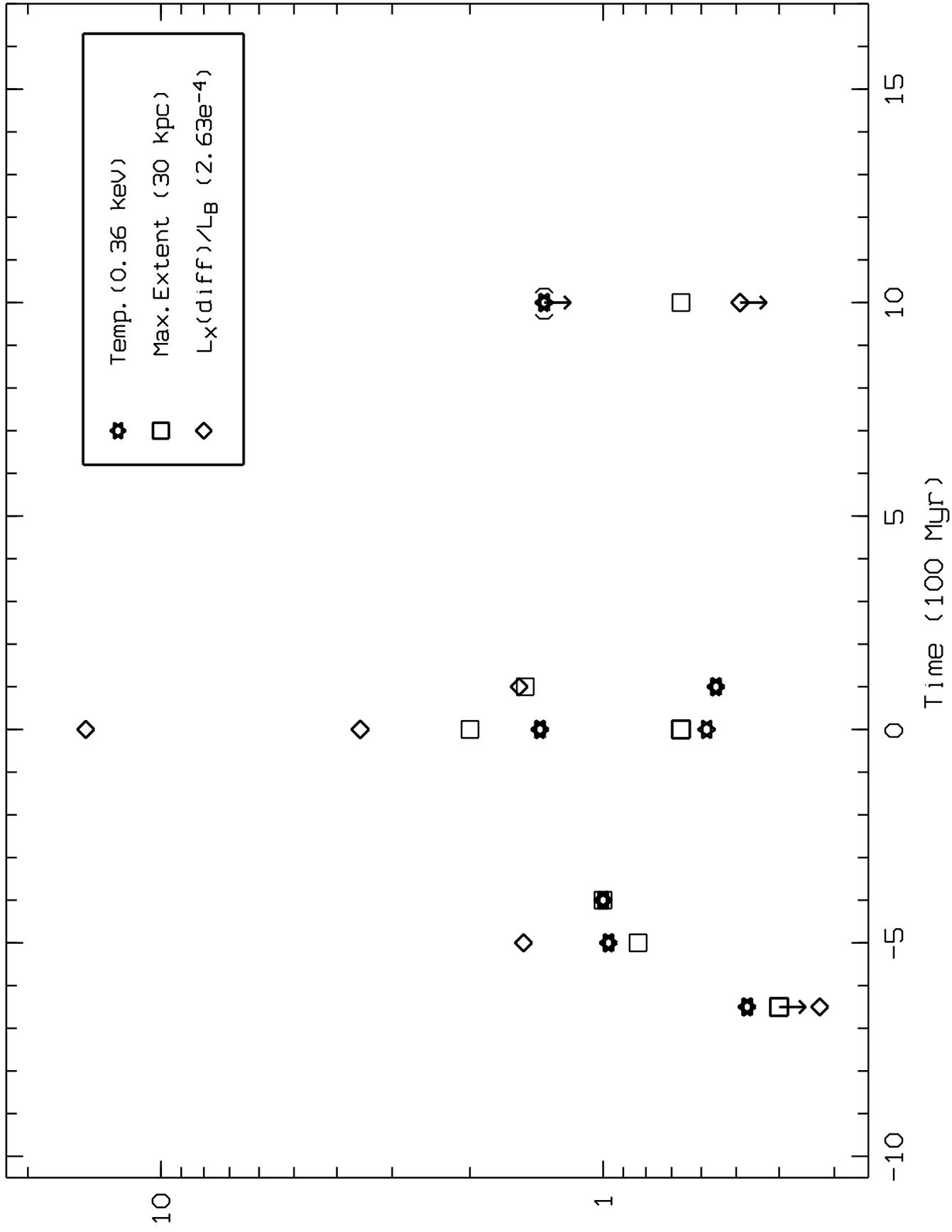
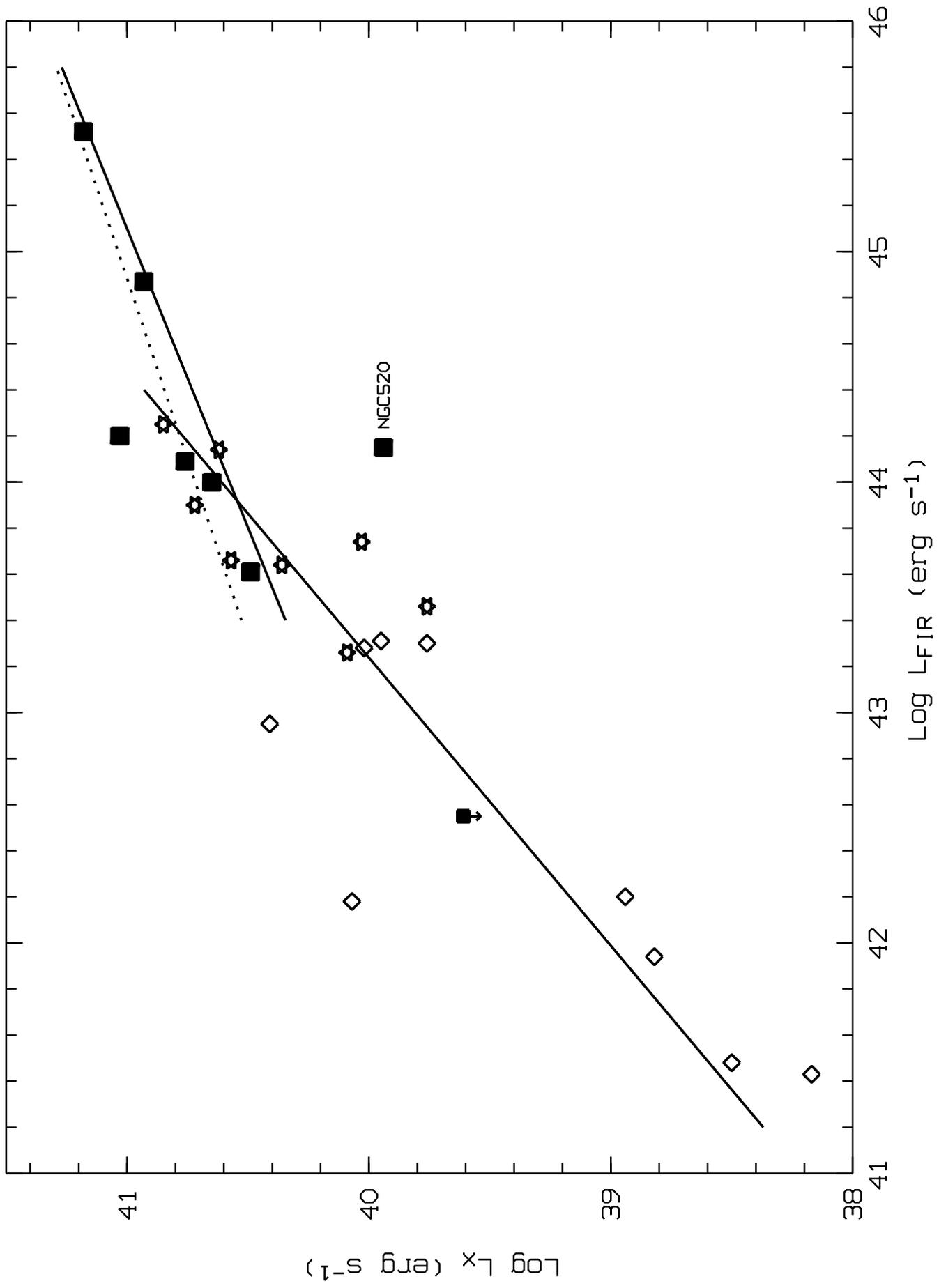
\end{document}